\documentclass[reprint,
 amsmath,amssymb,
 aps,prc,
 sort&compress,round,sort,numbers,
]{revtex4-2}

\usepackage{graphicx}
\usepackage{dcolumn}
\usepackage{bm}
\usepackage{comment,array}
\usepackage{natbib}
\usepackage{hyperref}
\usepackage[T1]{fontenc}

\begin{document}


\title{The nuclear symmetry energy from neutron skins and pure neutron matter in a Bayesian framework}

\author{William G. Newton}
\email{william.newton@tamuc.edu}
\author{Gabriel Crocombe}%
\affiliation{Department of Physics and Astronomy, Texas A\&M University-Commerce, Commerce, TX 75429-3011, USA}

\date{\today}

\begin{abstract}
We present an inference of the nuclear symmetry energy magnitude $J$, the slope $L$ and the curvature $K_{\rm sym}$ from combining neutron skin data on calcium, lead and tin isotopes and our best theoretical information about pure neutron matter. A Bayesian framework is used to consistently incorporate prior knowledge of the pure neutron matter equation of state from chiral effective field theory calculations. Neutron skins are modeled in a fully quantum Skyrme-Hartree-Fock approach using an extended Skyrme energy-density functional which allows for independent variation of $J$, $L$ and $K_{\rm sym}$ without affecting the symmetric nuclear matter equation of state. The effect of using neutron skin data obtained with different physical probes is quantified. We argue that, given the existing data, combining the errors in quadrature is the more appropriate way to obtain unified errors for each nuclide, and in doing so we obtain 95\% credible values of $J=31.3\substack{+4.2 \\ -5.9}$ MeV, $L=40\substack{+34  \\ -26}$ MeV and $K_{\tau} = L - 6K_{\rm sym}= -444\substack{+100 \\ -84}$ MeV using uninformative priors in $J$, $L$ and $K_{\rm sym}$, and $J=31.9\substack{+1.3 \\ -1.3}$ MeV, $L=37\substack{+9  \\ -8}$ MeV and $K_{\tau} = -480\substack{+25 \\ -26}$ MeV using pure neutron matter (PNM) priors. We also show that the non-positive correlation between $J$ and $L$ induced by neutron skin data is consistent with the nuclear droplet model. Neutron skin data alone is shown to place limits on the symmetry energy parameters as stringent as those obtained from chiral effective field theory alone, and when combined the 95\% credible intervals are reduced by a factor of 4-5. It is also shown that the majority of nuclear interactions used in the literature have sub-saturation density-dependencies that are inconsistent with the combination of PNM priors and neutron skin data. We show measurements of lead and calcium neutron skins from upcoming parity-violating electron scattering experiments at Jefferson Lab (PREX-II and CREX) and Mainz Superconducting Accelerator (MREX) should obtain total error ranges $\Delta L\approx50$ MeV and $\Delta K_{\tau} \approx 240$ MeV for uninformative priors and $\Delta L\approx30$ MeV and $\Delta K_{\tau} \approx 100$ MeV for PNM priors at 67\% credible bounds. Ahead of those experiments, we make predictions based on existing data on neutron skins of tin alone for the neutron skins of calcium and lead of 0.166$\pm$0.008 fm and $0.169 \pm 0.014$ fm respectively, using uninformative priors, and 0.167$\pm$0.008 fm and $0.172 \pm 0.015$ fm respectively, using PNM priors.
\end{abstract}

\maketitle


\section{\label{sec1}Introduction\protect}

Learning more about the behavior of the nuclear force in neutron-rich environments is a priority for nuclear structure and nuclear astrophysics. In particular, the crusts and outer cores of neutron stars are stabilized by essentially pure (superfluid) neutron matter (PNM). To make effective use of the data emerging from the burgeoning field of multi-messenger astronomy, neutron star modeling must incorporate our best knowledge of neutron matter. Neutron skins - defined as the difference between the root-mean-square radii of neutrons and protons in a nucleus $\Delta r_{\rm np} = (\langle \Delta r_{\rm n} \rangle^2)^{1/2} - (\langle \Delta r_{\rm p} \rangle^2)^{1/2}$ - are the most accessible neutron-rich environments on Earth and have been the subject of over four decades of experimental investigation. The neutron skins of calcium and lead are currently subject to a program of measurement independent of the strong interaction with all its obscuring complexity. These experiments use electrons as a weak probe - the parity-violating electron scattering method \citep{Roca-Maza:2011aa,Horowitz:2012aa} - at Jefferson lab (the PREX-II and CREX experiments) and at Mainz energy-recovering superconducting accelerator (the upcoming MREX experiment) \cite{Becker:2018aa,Thiel:2019aa}. Alongside this experimental work, the theoretical field of chiral-effective field theory (chiral-EFT) has allowed us to calculate the pure neutron matter (PNM) equation of state (EOS) with well defined and meaningful theoretical errors \cite{Gandolfi:2009qf,Gezerlis:2010fu,Hebeler:2010kb,Tews:2013yg,Gezerlis:2013lr,Kruger:2013fv,Gandolfi:2014qv,Gandolfi:2015nr,Sammarruca:2015lr,Tews:2016ty,Lynn:2016sf, Holt:2017lr}. In this paper we seek meaningful gains in our knowledge of neutron-rich matter by combining these two domains consistently \cite{Reinhard:2010aa}.

A powerful bridging concept between these two domains - the simpler abstraction of nuclear matter and the complex real-world nuclear system - is the symmetry energy, a quantity implicated in almost as many nuclear and neutron star observables as it has symbols in the literature; here we denote it as $S(\rho)$. The symmetry energy intuitively can be thought of as the energy requirement to turn symmetric nuclear matter (SNM) - equal number of neutrons and protons - into pure neutron matter (PNM). It is customary to parameterize the symmetry energy by its expansion in the density parameter $\chi = {(\rho - \rho_0) / 3\rho_0}$ around nuclear saturation $\rho_0=0.16$fm$^{-3}$;  $J$, $L$, and $K_{\rm sym}$ are the first three Taylor series coefficients, and are referred to simply as the magnitude, slope and curvature of the symmetry energy at saturation density:

\begin{equation}
S(\rho) = J + \chi L + {\frac{1}{2}} \chi^2 K_{\rm sym} + \dots
\end{equation}

Constraining the symmetry energy, and associated nuclear observables, has become a priority in the field of nuclear physics over the past two decades \cite{Tsang:2012qy,Li:2014fj,Lattimer:2014uq,Horowitz:2014aa,Tsang:2019ab}, and the strong links between the symmetry energy and neutron star properties provides additional motivation for measuring the size of neutron skins \cite{Brown:2000aa, Horowitz:2001aa, Fattoyev:2018aa} and determining the PNM EOS \cite{Gandolfi:2012lr,Hebeler:2010rz,Tews:2018aa}.

There have been many studies using existing nuclear energy-density functionals (EDFs) to derive symmetry energy constraints from neutron skin data and determine the impact of future measurements on symmetry energy constraints by examining correlations between the neutron skins and the symmetry energy parameters. Much care must be taken interpreting these results, however. Nuclear models contain many parameters, most of which are already fit to subsets of nuclear data, which induces correlations between the parameters of the energy density functional and thus nuclear matter parameters. 

Nuclear mass data has been shown to induce a positive correlation between $J$ and $L$, which because the data tends to fix the surface symmetry term, which is determined mainly by the symmetry energy at (2/3) $\rho_0$, $S(0.1$fm$^{-3}) \approx J - L/9$ (see, e.g. \citep{Lattimer:2013uu}). When one conducts a systematic review of all Skyrme models, a similarly positive correlation emerges for the same reason. One must then take care interpreting the correlations that emerge from further subjecting the models to neutron skin data, since they will be a convolution of the information content of neutron skin data and existing correlation induced by data that has already been applied to constrain the models used in the analysis. 

One approach to this is ``meta-modeling'' \cite{Margueron:2018aa,Margueron:2018ab, Traversi:2020tl} - systematically exploring the model space with respect to a number of parameters. The classic example of ``meta-modeling'' applied to neutron skin measurements is the analysis of \cite{Chen:2010aa}. In this work, tin-skins were modeled using Skyrme-Hartree-Fock (SHF) calculations of the Skyrme EDF. However, instead of using a disparate array of existing parameterizations of the Skyrme EDF, the authors fit a model to a minimal subset of nuclear data and then used two of the EDF parameters as handles to vary $J$ and $L$ independently while holding fixed the properties of symmetric nuclear matter. Applying a $\chi$-squared fit to the data, a 1$\sigma$ significance band was mapped out in the $J$ and $L$ plane in which a negative correlation between the two parameters was manifest. The neutron skin data was treated agnostically, with multiple independent measurements of a particular nuclide's skin combined by taking the overall highest and lowest values reported. This might not be the best way to treat the experimental data, however, as it can discount experimental studies with (possibly justifiably) smaller errors.  

Used in this way, the Skyrme EDF can be itself viewed as a ``meta-model.'' However, the traditional Skyrme model contains only enough degrees of freedom to vary $J$ and $L$ independently without disturbing SNM properties, there is an in-built relation (linear, in the Skyrme model) between the third symmetry energy parameter $K_{\rm sym}$ and $J$ and $L$. This could be a confounding factor in studying neutron skins, since the neutron skin has been shown to $K_{\rm sym}$ \cite{Raduta:2018aa} and the symmetry compressibility $K_{\tau}$ \cite{Centelles:2009aa}. 

Different definitions of $K_{\tau}$ exist in the literature, differing in the order of the density expansion of the EOS they take into account \cite{Chen:2009it}. Because previous constraints on $K_{\tau}$ obtained from nuclear resonances, heavy ion collisions and neutron skins \cite{Chen:2005aa,Shetty:2007aa,Li:2007aa,Centelles:2009aa} use the definition $K_{\tau} = K_{\rm sym} - 6L$, this is the one we shall use.

In the intervening decade much progress has been made constraining the PNM EOS \cite{Gandolfi:2009qf,Gezerlis:2010fu,Hebeler:2010kb,Tews:2013yg,Gezerlis:2013lr,Kruger:2013fv,Gandolfi:2014qv,Gandolfi:2015nr,Sammarruca:2015lr,Tews:2016ty,Lynn:2016sf, Holt:2017lr}. The slope of the symmetry energy $L$ is directly proportional to the pressure of PNM at saturation density, and hence it is particularly transparent in its physical connection with both neutron star radii and neutron skin thicknesses \cite{Fattoyev:2018aa}. A natural way of combining theoretical PNM EOS constraints with neutron skin constraints on the symmetry energy is to treat the PNM constraints as \emph{prior} knowledge of the symmetry energy ahead of an application of neutron skin data in a Bayesian probabilistic approach.

In this study we revisit and extend \citet{Chen:2010aa} it in the following ways. (1) We use a Bayesian inference approach (a) which allows us to explicitly and consistently incorporate prior nuclear matter knowledge, and is thus an appropriate framework to incorporate knowledge of the PNM EOS into the neutron skin analysis, and derive constraints on the PNM EOS with it, and (b) frees us from the $\chi$-squared requirement that the probability distributions be normal within the significance interval we are examining. (2) We use an extended Skyrme model that allows us to vary $K_{\rm sym}$ as well as $J$ and $L$ independently; our results will be posterior probability distributions over those parameters and any relevant ones derived from them, and (3) we will examine the effect of neutron skin data selection and combination on the results obtained. As well as tin isotopes, we will use neutron skin data of $^{48}$Ca and $^{208}$Pb.

In section II we outline our Skyrme-Hartree-Fock Approach to modeling neutron skins, in section III we describe the Bayesian framework including a discussion of our prior probability distributions over the parameters $J$, $L$ and $K_{\rm sym}$, in section IV we present our results, in section V we discuss more generally the correlations that arise between $J$ and $L$ in particular, and in section VI we place our results in context and give our conclusions. 

\section{Modeling Neutron Skins\protect}

We use the Skyrme Energy Density Functional $\mathcal{H}_{\rm Skyrme}$, \cite{Vautherin:1972aa,Bender:2003aa,Lim:2017aa} modified to include a less restrictive density dependence. It is composed of the following zero range, density-dependent, finite-range, gradient, spin-orbit, spin-gradient and Coulomb terms:

\begin{equation}
    \mathcal{H}_{\rm \delta} =  \frac{1}{4} t_{0} \rho^2 [(2+x_{0}) - (2x_{0}+1)(y_{p}^{2}+y_{n}^{2})] 
\end{equation}

\begin{align}
    \mathcal{H}_{\rho} &=  \frac{1}{4} t_{3} \rho^{2+\alpha_3} [(2+x_{3}) - (2x_{3}+1)(y_{p}^{2}+y_{n}^{2})] \notag \\
    &+  \frac{1}{4} t_{4} \rho^{2+\alpha_4} [(2+x_{4}) - (2x_{4}+1)(y_{p}^{2}+y_{n}^{2})] 
\end{align}

\begin{align}
    \mathcal{H}_{\rm eff} &= \frac{1}{8} \rho [t_{1}(2+x_{1})+t_{2}(2+x_{2})]\tau
\notag \\  & + \frac{1}{8} \rho [t_{1}(2x_{1}+1)+t_{2}(2x_{2}+1)](\tau_{p}y_{p}+\tau_{n}y_{n})
\end{align}

\begin{align}
    \mathcal{H}_{\rm grad} &= \frac{1}{32} (\nabla \rho)^2 [3 t_{1}(2+x_{1})-t_{2}(2+x_{2})]
\notag \\  & - \frac{1}{32}  [3t_{1}(2x_{1}+1)+t_{2}(2x_{2}+1)][(\nabla \rho_{p})^2 + (\nabla \rho_{n})^2)
\end{align}

\begin{align}
    \mathcal{H}_{\rm so} &= {\frac{W_0}{2}} (\vec{\nabla} \rho \cdot \vec{J} + \vec{\nabla} \rho_{\rm p} \cdot \vec{J}_{\rm p} + \vec{\nabla} \rho_{\rm n} \cdot \vec{J}_{\rm n})
\end{align}

\begin{align}
    \mathcal{H}_{\rm sg} &= -{\frac{1}{16}} (t_1 x_1 + t_2 x_2) \vec{J}^2 + {\frac{1}{16}} (t_1 - t_2) [\vec{J}_{\rm p}^2 + \vec{J}_{\rm p}^2]
\end{align}

\begin{align}
    \mathcal{H}_{\rm Coul} (r) &= {\frac{1}{2}} e^2 \rho_{\rm p}(r) \int  \frac{{\rho_{\rm p} (r^{\prime}) dr^{\prime} } {|\vec{r} - \vec{r}^{\prime} |}} - {\frac{3}{4}} e^2 \rho_{\rm p} (r) \bigg( \frac{3 \rho_{\rm p} (r)}{\pi} \bigg)^{1/3}
\end{align}

\noindent where $\rho_i$, $\tau_i$ and $\vec{J}_i$ ($i=p,n$) are the density, kinetic energy density and spin-density respectively.

The most widely used version of the Skyrme EDF functional contains nine parameters $x_{0-3}, t_{0-3} $ and $\alpha_3$ that determine the nuclear matter EOS. However it doesn't contain sufficient degrees of freedom to vary the first three coefficients of the symmetry energy expansion independently while holding the SNM EOS constant. We therefore extend the density dependence of the Skyrme. We do this by adding a second density dependent term to $\mathcal{H}_{\rho}$ that is parameterized by $t_4$, $x_4$ and $\alpha_4$ \cite{Lim:2017aa}. There are a number of other ways of extending the Skyrme EDF \cite{Agrawal:2006aa,Erler:2010aa,Zhang:2016ww}; this is the simplest modification that has been explicitly shown to allow the Skyrme EDF to accurately describe the density dependence of pure neutron matter at low densities as predicted by chiral EFT calculations, and is thus an appropriate model to incorporate that \emph{ab-initio} information. As our baseline model, we use the Sk$\chi$450 parameter set from  Table~1 of \cite{Lim:2017aa}, fit to the properties of doubly magic nuclei and, importantly, chiral-EFT numerical data.

One can invert the resulting equations to find, in particular, an expression for the Skyrme parameters $x_0$, $x_3$ and $x_4$ in terms of the symmetry energy parameters $J$, $L$ and $K_{\rm sym}$. This inversion is inevitably more complicated with the additional parameters, but can nevertheless be written analytically as follows 

\begin{equation} 
x_0 = 1 - {8D_0 \over t_0}; \;\;\; x_3 = 1 - {16D_3 \over t_3}; \;\;\;
x_4 = 1 - {16D_4 \over t_4}
\end{equation}

\noindent where:

\begin{align*} 
&D_0 = [9J^\prime{(\alpha_3 + 1)(\alpha_4 + 1)} - \notag \\
&\;\;\;\;\;\;\;\;\;\; 3L^\prime(\alpha_3 + \alpha_4 + 1) + K_{\rm sym}^\prime]9\alpha_3\alpha_4n_0 \notag \\
&D_3 = [(9J^\prime - 3L^\prime)(\alpha_4 + 1) + K_{\rm sym}^\prime]9(\alpha_3^2 - \alpha_3\alpha_4)n_0^{(\alpha_3 +1)} \notag \\
&D_4 = [(9J^\prime - 3L^\prime)(\alpha_3 + 1) + K_{\rm sym}^\prime]9(\alpha_4^2 - \alpha_3\alpha_4)n_0^{(\alpha_4 +1)}
\end{align*}

\noindent and

\begin{align} 
J^\prime &= J - D_{\rm KE}n_0^{2/3} - D_{\rm 12}n_0^{5/3} \notag \\
L^\prime &= L - 2D_{\rm KE}n_0^{2/3} + 5D_{\rm 12}n_0^{5/3} \notag \\
K_{\rm sym}^\prime &= K_{\rm sym} + 2D_{\rm KE}n_0^2/3 - 10D_{\rm 12}n_0^5/3 \notag \\
\end{align}

\noindent and

\begin{align}
D_{\rm KE} &= {\hbar^2 \over 12m}\bigg({3\pi^2 \over 2}\bigg)^{2/3} \notag \\
D_{\rm 12} &= {2 \over 3}\bigg({3\pi^2 \over 2}\bigg)^{2/3}{1 \over 16}\bigg(-3t_1x_1 + 5t_2x_2 + 4t_2\bigg)
\end{align}

\noindent This enables us obtain a unique Skyrme model characterized by different values of $x_0$, $x_3$ and $x_4$ for any value of $J$, $L$ and $K_{\rm sym}$ we choose. These are a specific form of the equations in appendix A of \citep{Zhang:2016ww}.

\section{Bayesian Inference of Model Parameters from Data}

In a Bayesian framework, our goal is to estimate the probability distribution of Skyrme models - characterized by unique values of the parameters $J$, $L$ and $K_{\rm sym}$ - as inferred from neutron skin data $\mathcal{D}$:

\begin{align}
P(J,L,K_{\rm sym}|&\mathcal{D}) \notag  \\
&= {1 \over N} P(\mathcal{D}|J,L,K_{\rm sym})P(J,L,K_{\rm sym}) 
\end{align}

\noindent where $P(\mathcal{D}|J,L,K_{\rm sym})$ is the likelihood function, $P(J,L,K_{\rm sym})$ is the prior probability distribution on the parameters and $N$ is a normalization factor. The priors are an important facet of Bayesian probability, as they allows us to consistently incorporate our prior knowledge of model parameters into our analysis. For example, if we want to express the fact that we know nothing about $J$, $L$ and $K_{\rm sym}$ within a given range, then we would set $P(J,L,K_{\rm sym})=$constant - i.e. each value is equally likely to start with. These are called uniform or uninformative priors.  

The likelihood function can be written

\begin{align}
P(\mathcal{D}|&J,L,K_{\rm sym}) \notag \\ = &\int  P(\mathcal{D}|\Delta r_{\rm np}(J,L,K_{\rm sym})) \times \\ \notag
&P(\Delta r_{\rm np}(J,L,K_{\rm sym})|J,L,K_{\rm sym}) d \Delta r_{\rm np}
\end{align}

\noindent where $P(\Delta r_{\rm np} (J,L,K_{\rm sym}) |J,L,K_{\rm sym})$ is the distribution of our model predictions - the distribution of neutron skin values we get out of our Skyrme-Hartree-Fock (SHF) calculations given our range of input values of $J$, $L$ and $K_{\rm sym}$. Assuming Gaussian errors on the data, the probability distribution of the data given a value of the neutron skin is given by

\begin{align}
P(\mathcal{D}|&\Delta r_{\rm np}(J,L,K_{\rm sym})) = \\ \notag & \exp{\bigg[- {(\Delta r_{\rm np}^{\rm data} - \Delta r_{\rm np}(J,L,K_{\rm sym}))^2 \over \sigma^2} \bigg]}
\end{align}

\noindent where $\sigma$ are the $1\sigma$ errors reported in the literature. Finally, a note about terminology: in a Bayesian framework, posterior probability distributions are characterized by credible intervals for the model parameters - the random variables whose probability distributions we are inferring - given the data at hand, rather than confidence intervals within which we might expect to find the ``true'' model parameter.

\begin{figure*}[!t]
\includegraphics[scale=0.52]{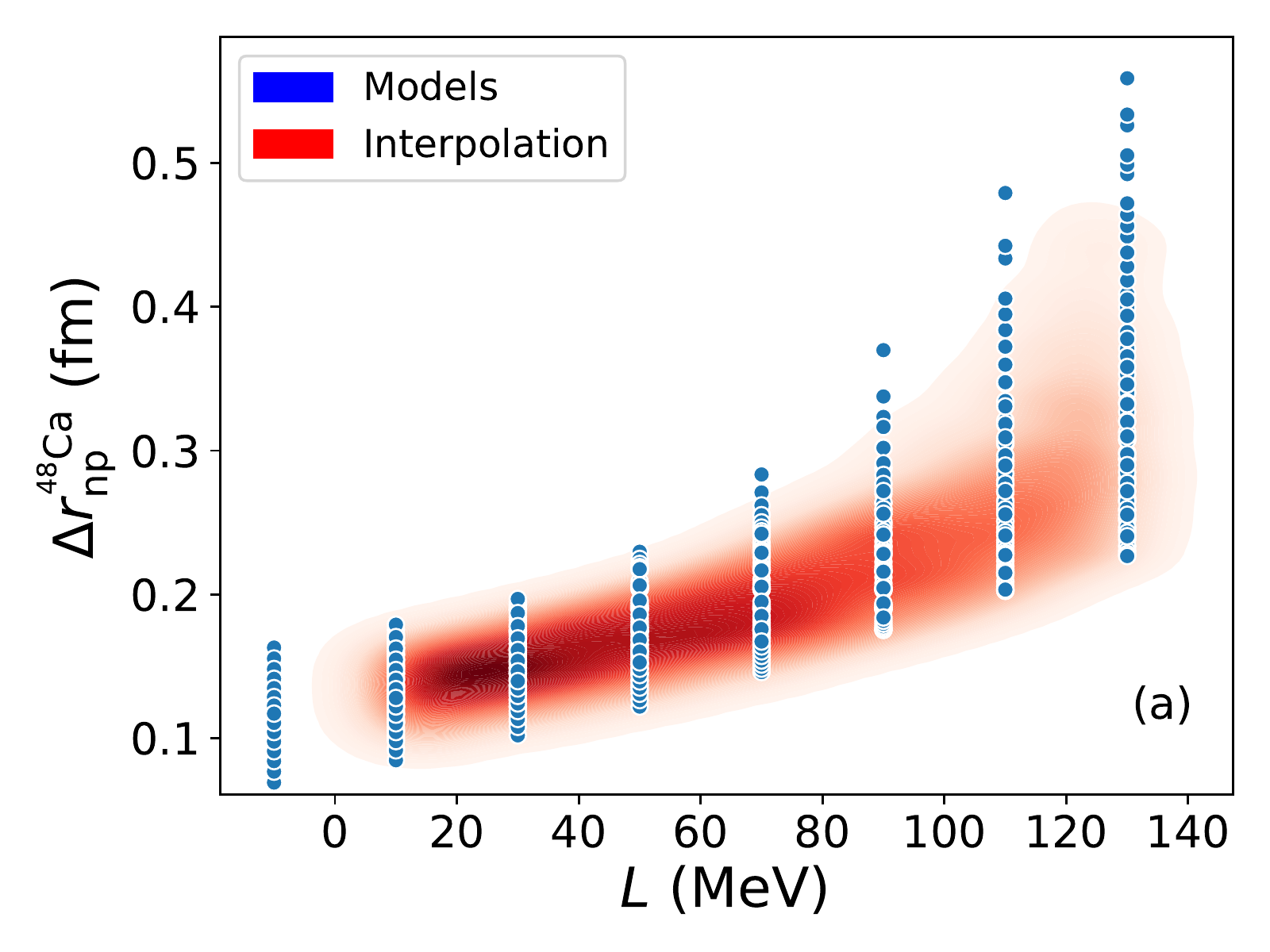}\includegraphics[scale=0.52]{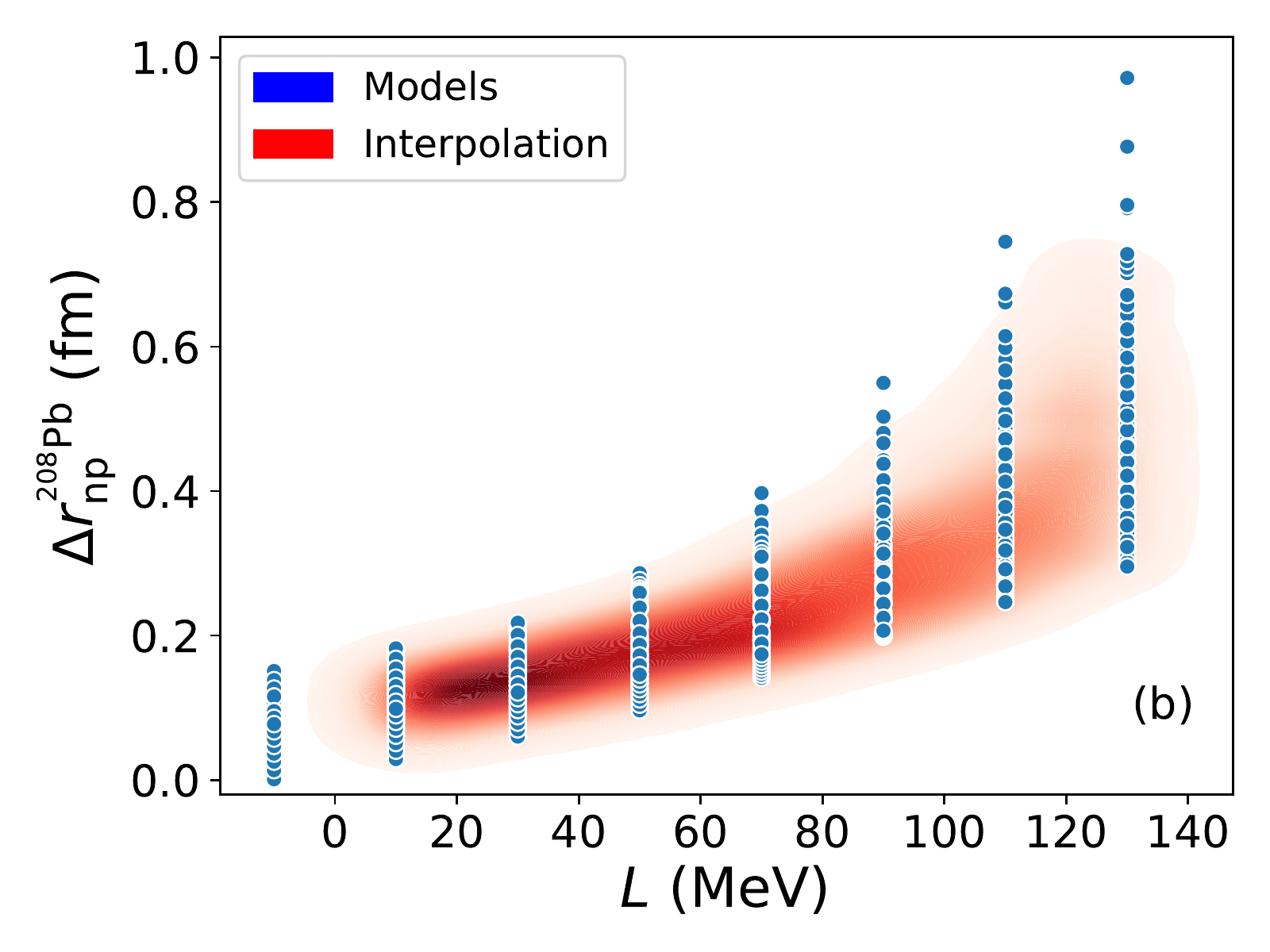}
\includegraphics[scale=0.52]{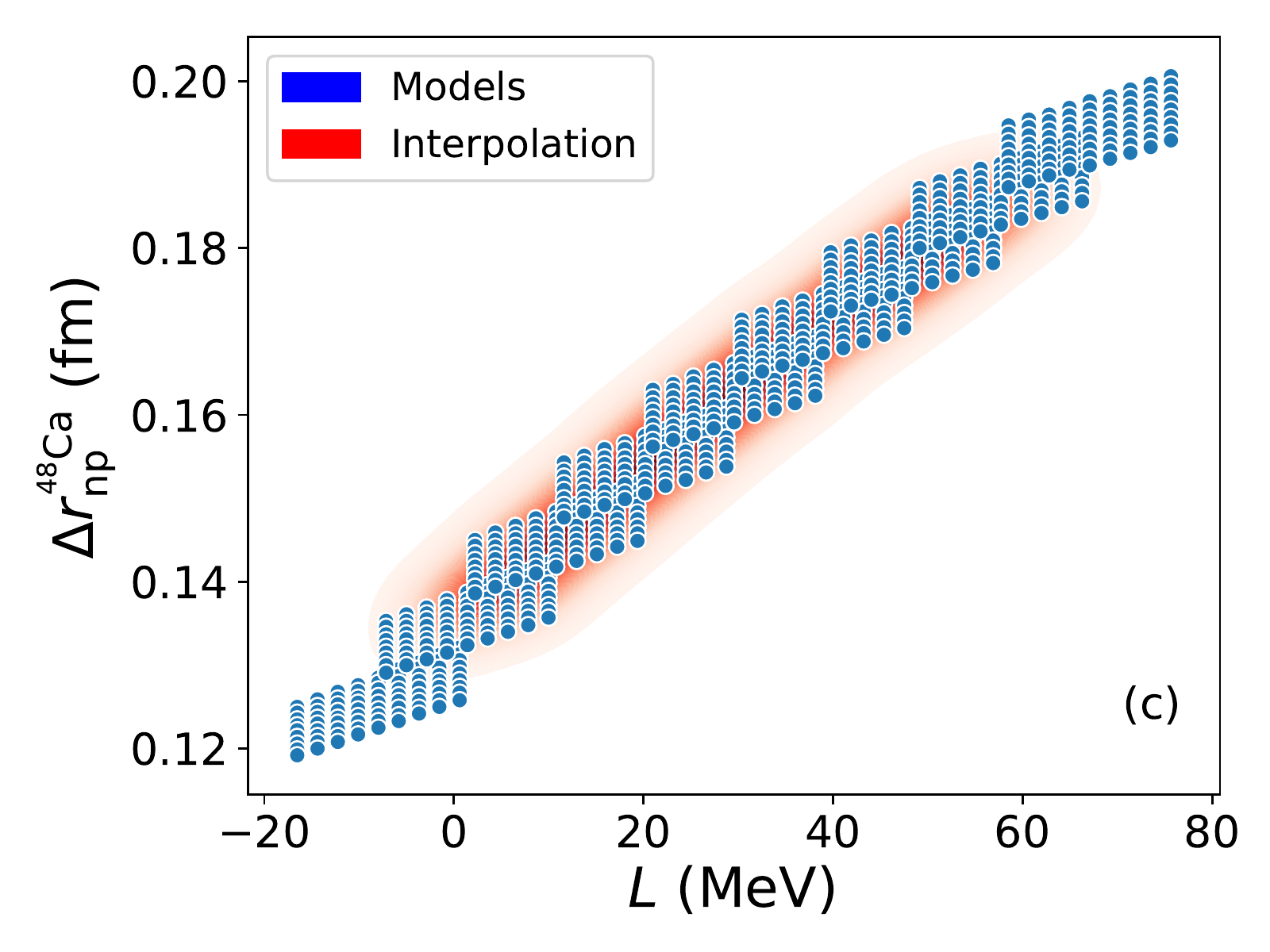}\includegraphics[scale=0.52]{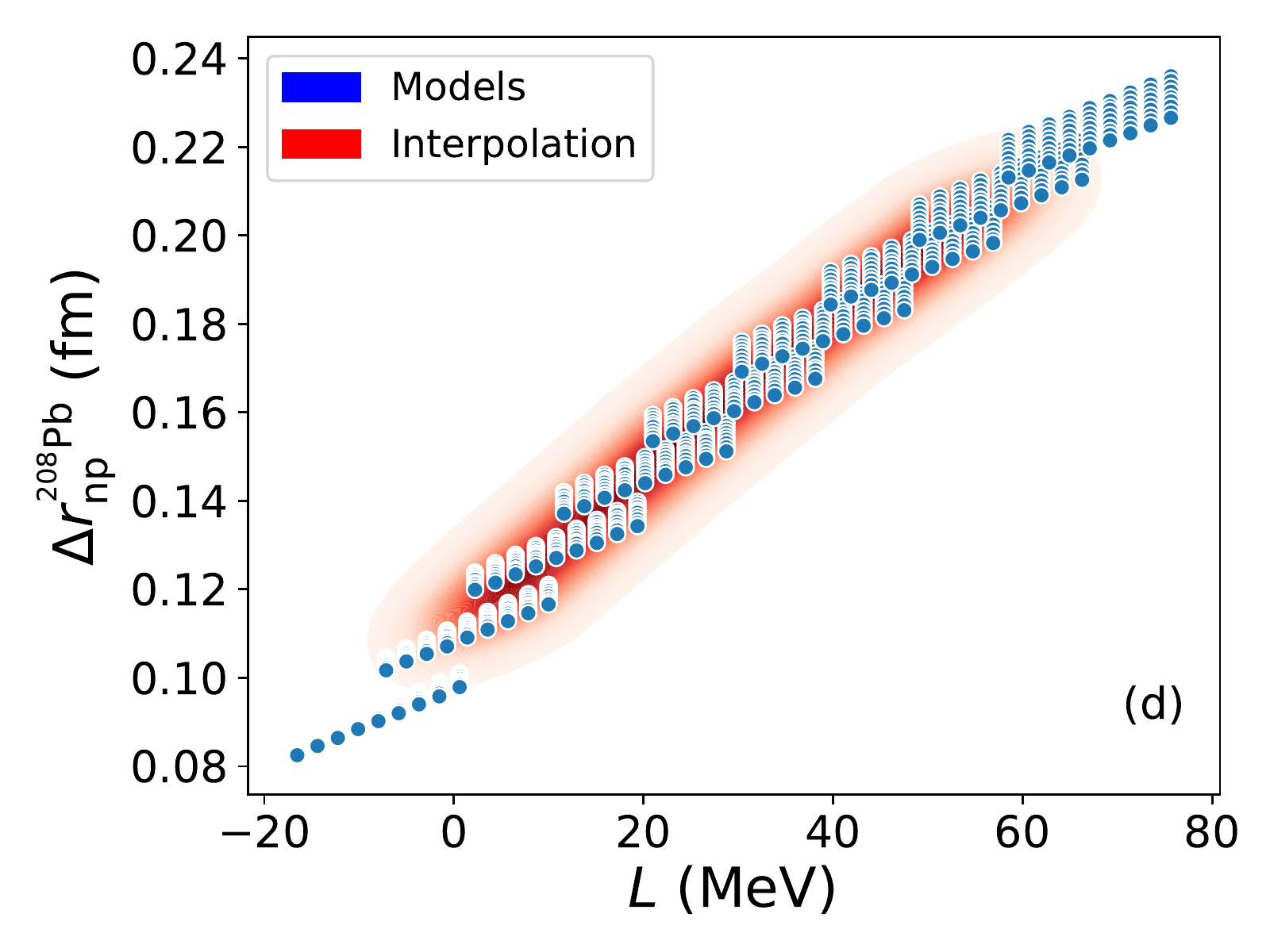}
\caption{Predictions for the neutron skins of $^{48}$Ca (a,c) and $^{208}$Pb (b,d) as calculated with our Skyrme-Hartreee-Fock models (blue points) and their interpolation (red density plot). Our models draw symmetry energy parameters from our uninformative prior distribution (a,b) and our pure neutron matter priors (c,d). A uninformative sampling of the symmetry energy parameters $J$, $L$ and $K_{\rm sym}$ leads to only a weak correlation between the neutron skin and $L$} \label{Fig:1}
\end{figure*}

\begin{figure*}[!t]
\includegraphics[scale=0.37]{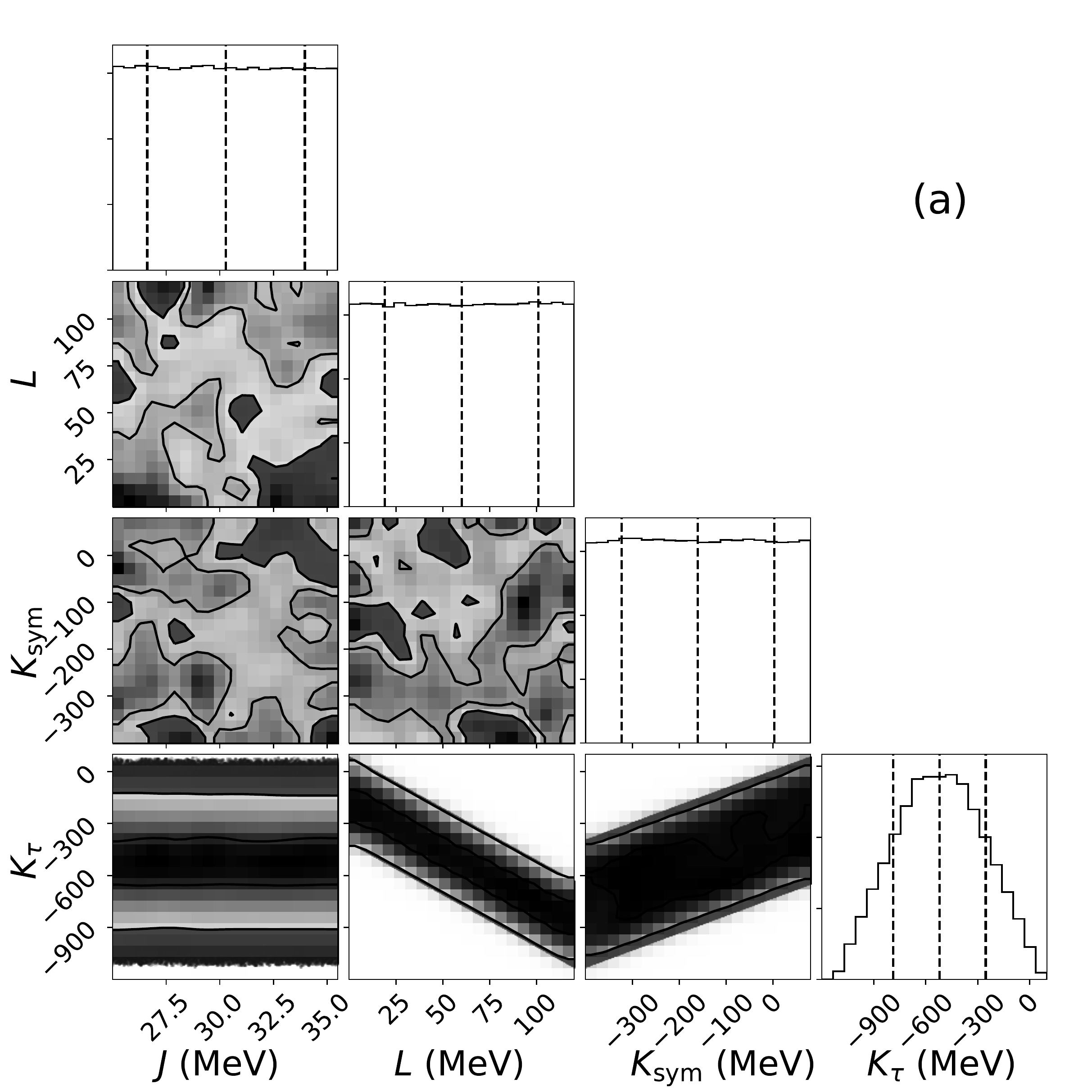}\includegraphics[scale=0.37]{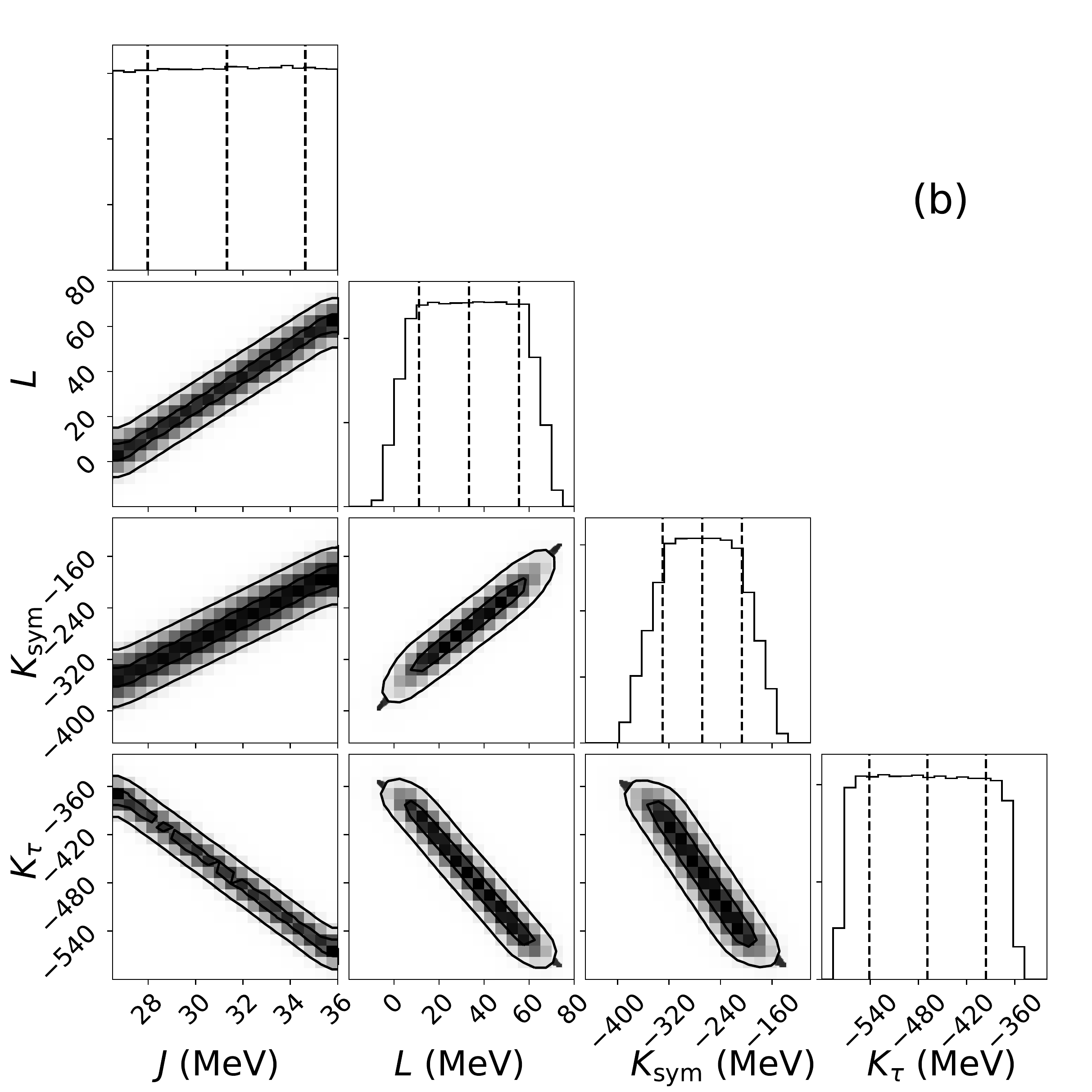}\
\caption{Two-dimensional and one-dimensional marginalized distributions of our prior distributions of symmetry energy parameters $J$, $L$ and $K_{\rm sym}$ as well as the symmetry compressibility $K_{\tau}$, in units of MeV. We show our uninformative prior (a), and on the right our pure neutron matter priors (b) by setting the data errors in our simulation to be much larger than the data values; the first three two-dimensional distributions in (a) are uninformative, the small visible fluctuations being statistical.} \label{Fig:2}
\end{figure*}

\subsection{Priors}

We directly parameterize our Skyrme models using the first three coefficients in the density expansion of the symmetry energy: $J$ (the value of $S(\rho)$ at saturation density); $L$ (the slope of $S(\rho)$ at saturation density); and $K_{\rm sym}$ (the curvature at saturation density). As part of the Bayesian framework, we must explicitly include our prior probabilities for the distributions of these parameters. We will build our analysis around two different priors.

The first prior will be a uniform distribution over a conservative range $J$=24 to 36 MeV, $L$=-10 to 130 MeV and $K_{\rm sym}$ = -440 to 120 MeV. These are referred to as uninformative priors. The limits of the range are still an explicit statement that the true value of the symmetry energy parameters cannot be outside these ranges; this is based upon over two decades of activity to constrain the density dependence of the symmetry energy by confronting models with experimental data \cite{Tsang:2012qy,Li:2014fj,Lattimer:2014uq,Horowitz:2014aa,Tsang:2019ab}.

For the second prior we choose to use our theoretical knowledge of pure neutron matter (PNM) from Chiral EFT computations. These constrain $J$, $L$ and $K_{\rm sym}$. We want to choose our prior carefully, so that we incorporate the knowledge gained in these calculations but acknowledge their existing uncertainties, for example in the order-by-order convergence of the chiral-EFT models. A useful way to parameterize these models is through a Taylor expansion of the Fermi liquid parameters that characterize the two-neutron interaction energy \cite{Holt:2018aa,Holt:2018ab}. In particular, the symmetry energy at a given density is related to 3$f_0'$ - $f_1$ where $f_0'$ is the isotropic, isovector Fermi liquid parameter and $f_1$ the second isoscalar Fermi liquid parameter; expanding 3$f_0'$ - $f_1$ about a reference density gives expansion coefficients $a_i$. We use the two parameters $a_0$ and $b_{\rm 12}=\eta_1(a_2-a_1)$ where $\eta_1$ is a parameter related to the reference density (see \cite{Holt:2018aa} for details). 

Chiral EFT allows us to constrain these parameters to the conservative range $J$=24-36MeV, $a_0$ = 5.53-6.41 fm$^2$ and $b_{\rm 12}$= 0 to 16 fm$^2$ \cite{Holt:2018aa}. Our PNM priors are drawn uniformly from these three ranges and then translated into distributions of $L$ and $K_{\rm sym}$ using the relations \cite{Holt:2018aa}:

\begin{equation}
    L = 6.7J + C_{L},
\end{equation}
\noindent and calculate $K_{\rm sym}$ by:
\begin{equation}
    K_{\rm sym} = 18.4J + C_{K_{\rm sym}},
\end{equation} 
\noindent where $C_{L}$ and $C_{K_{\rm sym}}$ are given respectively as:

\begin{equation}
    C_{L} = -19.47 a_{0} + 1.56 b_{12}-59.22
\end{equation}

\begin{equation}
    C_{K_{\rm sym}} = 5(C_{L}+50.22)+7.79 b_{12} -258.3 .
\end{equation}

Each value of $J$, $L$ and $K_{\rm sym}$ drawn from our priors corresponds to a different Skyrme model. We then calculate the neutron skins using the Skyrme-Hartree-Fock (SHF) code \texttt{Sky3d} \cite{Maruhn:2014aa} modified to include the extra density-dependent terms in the Hamiltonian.

We perform Skyrme-Hartree-Fock calculations for Skyrme models over a 9x9x9 grid of $J$, $L$ and $K_{\rm sym}$ points, which gives us sufficiently dense coverage of parameter space to interpolate the neutron skins of nuclides accurately at points in between. This is how we efficiently sample $\sim 10^6$ points from our functions $P(\Delta r_{\rm np}(J,L,K_{\rm sym})|J,L,K_{\rm sym})$ in order to perform our Bayesian analysis. Our interpolation reproduces the calculated values of neutron skins to an average of one part in $10^{7}$. In Figure~1, we plot the neutron skins from our SHF calculations as a function of the density dependence of the symmetry energy $L$ for $^{48}$Ca (left) and $^{208}$Pb (right) nuclides and for our uninformative (top) and PNM (bottom) priors. These are overlaid on a density plot of $\sim 10^5$ points obtained using our interpolating functions. Although the density contours do not extend right to the boundaries of our model parameters, values are being sampled there, just not in sufficient numbers to be visible. Note that, particularly over the uninformative priors, the correlation of the neutron skins with $L$ is not strong (especially at higher values of $L$). Imposing the PNM prior strengthens the correlation considerably. Reported strong correlations between neutron skins and $L$ tend are influenced by correlations induced by models used being already fit to other subsets of nuclear data, or models which by construction have such correlations in-built (such as minimal Skyrme models). Finally, we note that the mean absolute deviation of our model predictions for the binding energies and charge radii of doubly magic nuclei over the whole of our prior ensembles is 3\% and 1\% respectively.

\begin{figure*}[!t]
\includegraphics[scale=0.55]{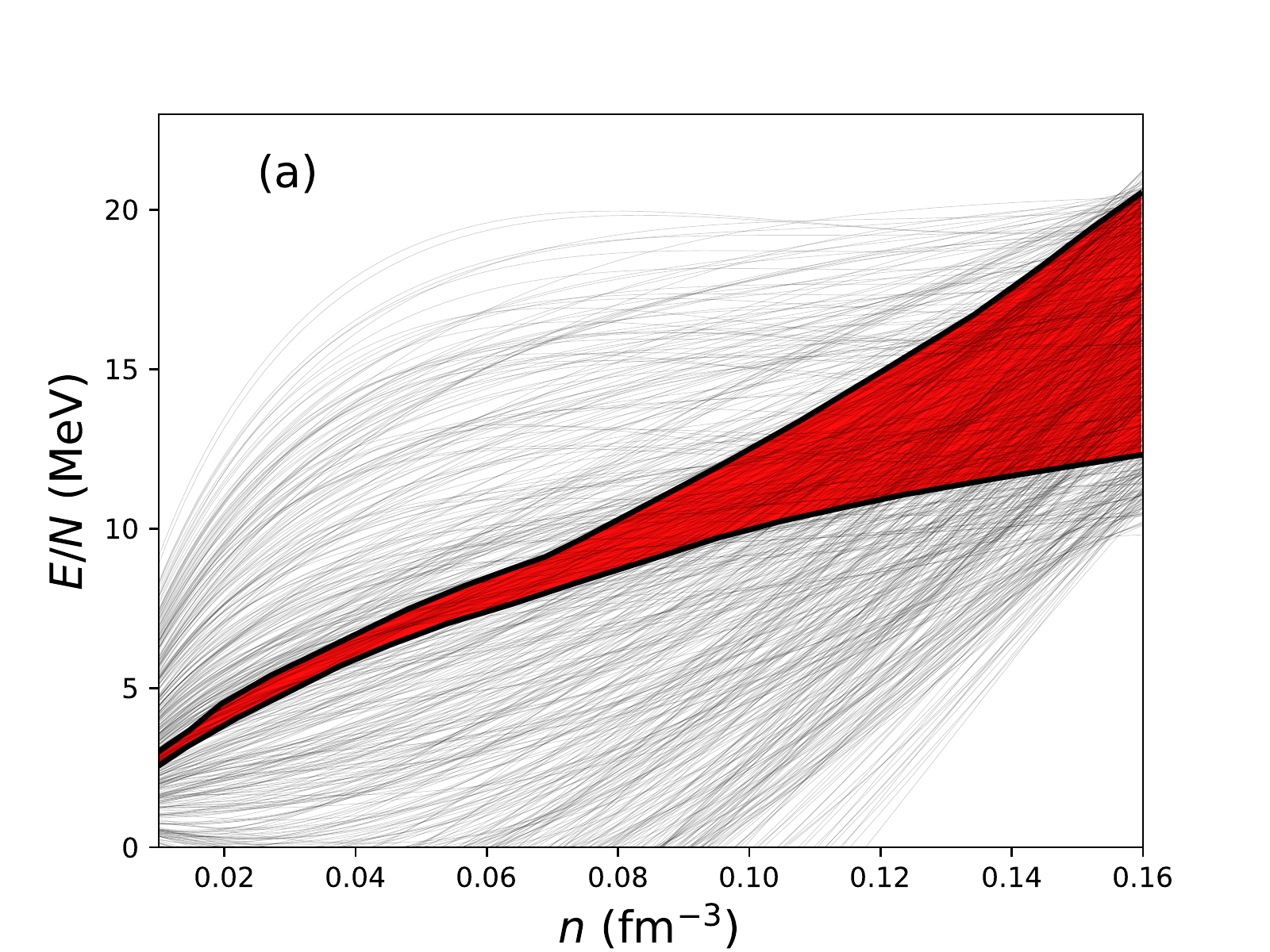}\includegraphics[scale=0.55]{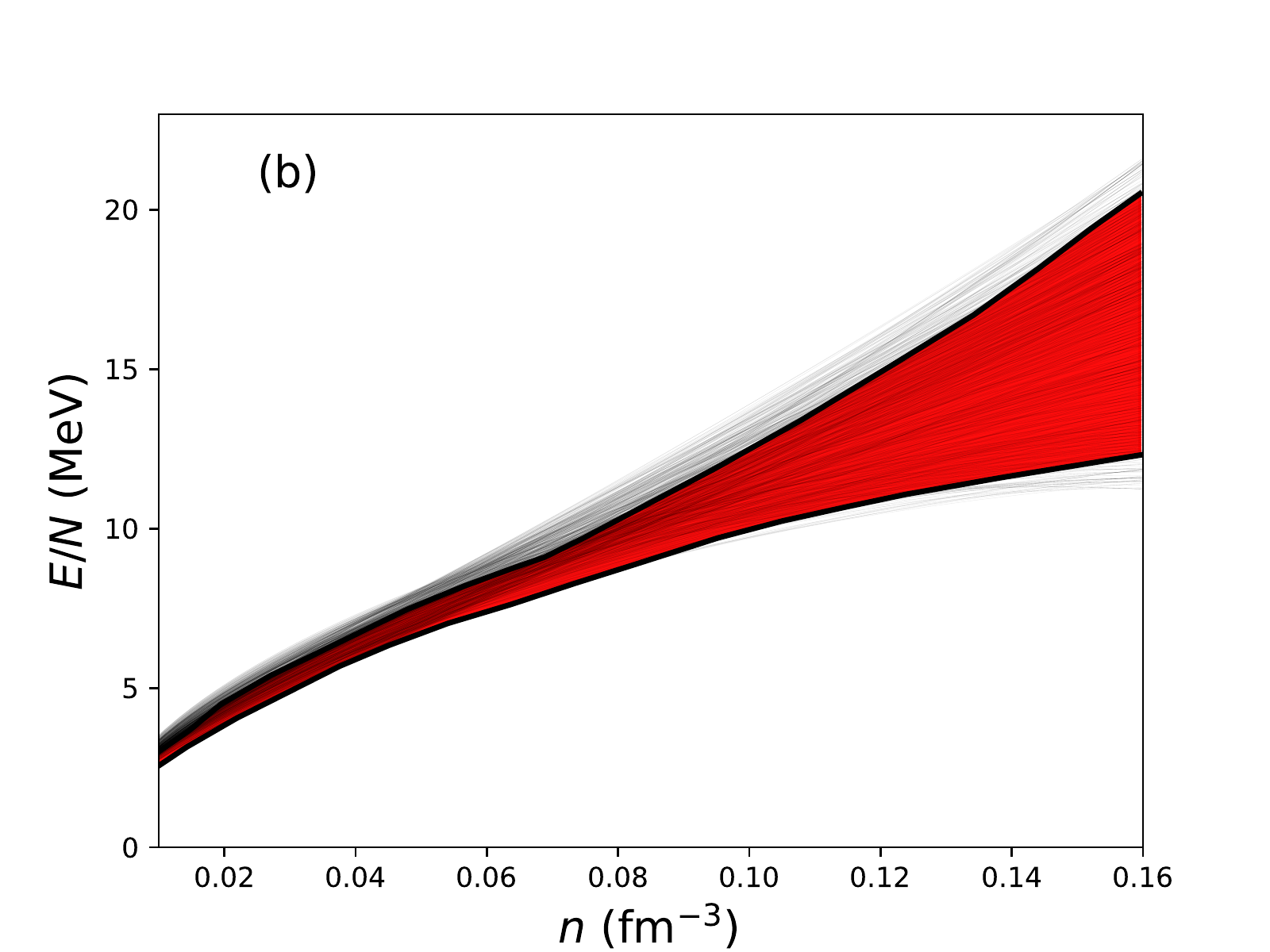}
\caption{Several hundred pure neutron matter equations of state sampled from our uninformative priors (a) and PNM priors (b) as a function of baryon density $n$ compared to a band extracted from a sample of $ab$-$initio$ calculations of the PNM EOS taken from \cite{Tews:2016ty}.} \label{Fig:3}
\end{figure*}

We perform the integration (equation~14) using Markov-Chain Monte-Carlo (MCMC) simulations using the \texttt{emcee} package \cite{Foreman-Mackey:2013aa}. In order to illustrate our prior distributions of $J,L$ and $K_{\rm sym}$, and test our MCMC simulations, we conducted simulations in the limit of very large data errors, to reproduce the priors. We show the results in Figure~2. We also include the resulting distribution of the symmetry compressibility $K_{\tau}$. One can see the uninformative priors manifest in Fig~2a, and as expected the $K_{\tau}$ priors show a negative correlation with $L$ and positive with $K_{\rm sym}$ by construction $K_{\tau} = K_{\rm sym} - 6 L$. The symmetry energy at sub-saturation densities - here represented by $\rho=0.1$fm$^{-3}$ - correlates positively with $J$ and negatively with $L$ (a steeper slope at saturation density will lead to a more rapid decline of $S(\rho)$ with density and hence a smaller density) and a slight correlation with $K_{\rm sym}$. The PNM priors already correlate $J$, $L$ and $K_{\rm sym}$ positively. Notably, this means the previous positive correlations involving $S(0.1$fm$^{-3})$ and $K_{\tau}$ are reversed by $L$'s positive correlation with $J$ and $K_{\rm sym}$.

We report our errors as 95\% credible intervals. We sample the interpolated data of order $10^6$ times in order that the 95\% credible intervals are adequately stable.

In Figure~3 we plot the EOS of PNM for models drawn from the uninformative priors (left) and from PNM priors (right). The red band is the region constrained by Chiral EFT \cite{Gandolfi:2009qf,Gezerlis:2010fu,Hebeler:2010kb,Tews:2013yg,Gezerlis:2013lr,Kruger:2013fv,Gandolfi:2014qv,Gandolfi:2015nr,Sammarruca:2015lr,Tews:2016ty,Lynn:2016sf, Holt:2017lr}. One can see that the PNM priors do indeed follow closely the predicted PNM band, whereas the uninformative priors allow us to explore a much wider range of possible EOSs.

\subsection{Data}

We subject our models to inferences of neutron skins from data gathered on the nuclides $^{48}$Ca, $^{112,114,116,118,120,122,124,130,132}$Sn and $^{208}$Pb. Here we review the different physical origins of neutron skin data sources. 

\subsubsection{Exotic atoms}

Studying the properties of atoms containing antiprotons and pions has been a fruitful source of data. After $\bar{p}-p$ and $\bar{p}-n$ annihilation events on the nucleus resulting from an incident beam of $\sim 100$ MeV antiprotons on the target, the ratio of $Z-1$ to $N-1$ products is sensitive to the surface proton and neutron distributions \cite{Trzcinska:2001aa,Trzcinska:2001ab}. This method is supplemented by X-ray spectroscopy determining the nuclear level shifts in anti-protonic atoms \cite{Trzcinska:2001aa,Trzcinska:2001ab,Kos:2007aa}. In order to extract the density distributions from the data, a parameterized function can be used such as a 2-parameter Fermi function \cite{Trzcinska:2001aa,Trzcinska:2001ab,Kos:2007aa,Brown:2007aa}, or density distributions calculated from microscopic models can be used \cite{Brown:2007aa}.

Additionally, the strong interaction between pions and the nucleus in pionic atoms \cite{Friedman:2007aa,Friedman:2012aa} and in pion scattering \cite{Friedman:2012aa} is sensitive to the nuclear density distribution and has been used to infer the neutron skin. 

\begin{figure*}[!ht]
\includegraphics[scale=0.7]{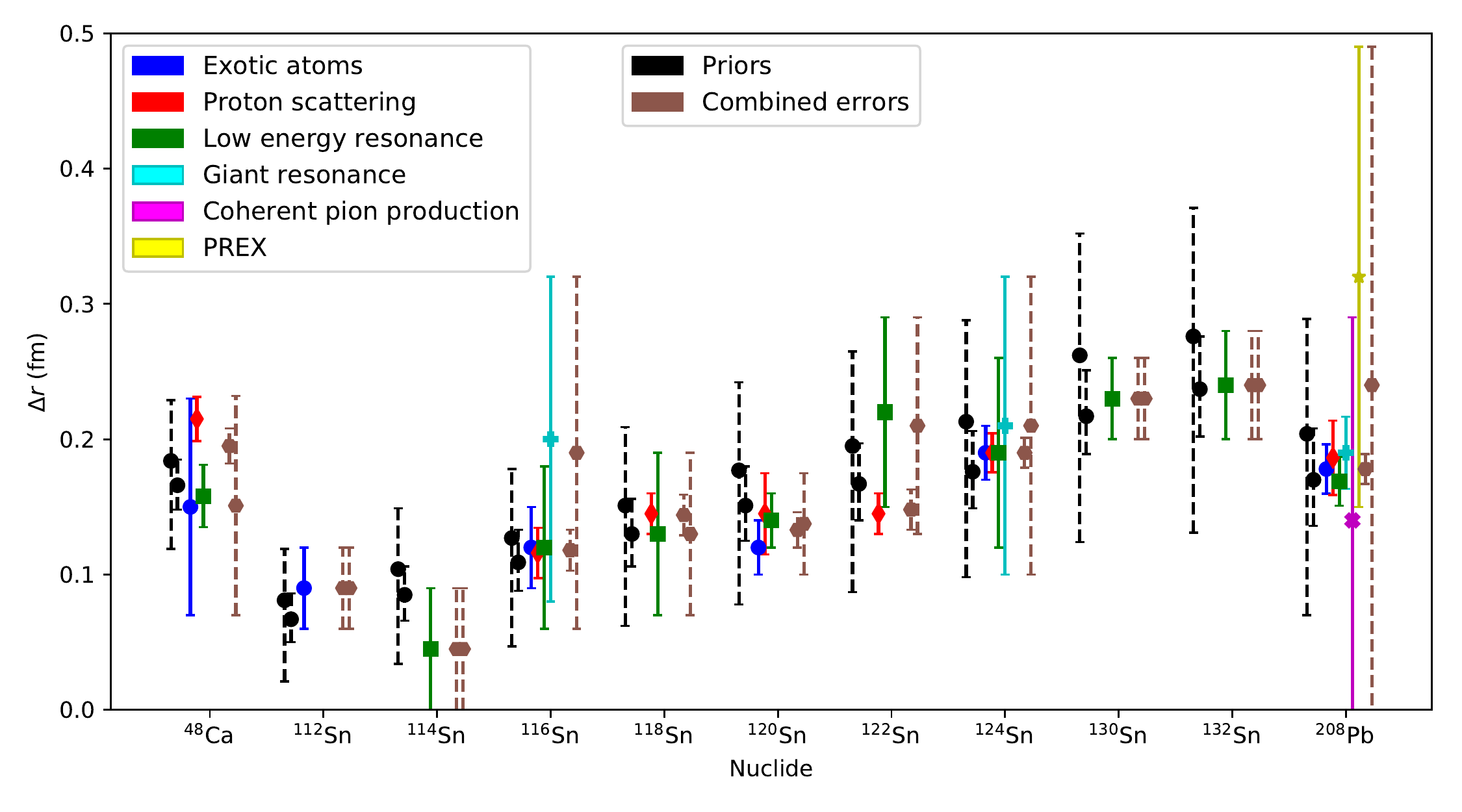}
\caption{Inferred values of neutron skins from different data sources (the solid, colored error bars) for each nuclide. These are compared with the 67\% credible ranges from our priors (the black dashed error bars) and the result of combining the data points (the brown dashed lines) as a total range (the larger of the combined errorbars) and in quadrature (the smaller of the combined errorbars).} \label{Fig:4}
\end{figure*}

There is good consistency between the nucleon density distributions obtained using antiprotonic and pionic atom data \cite{Friedman:2009aa}.

\subsubsection{Scattering}

The density distributions in nuclei have been studied using proton elastic scattering at a number of energies. At intermediate energies of 295MeV, the neutron skin of $^{48}$Ca \cite{Zenihiro:2018aa} and $^{208}$Pb \cite{Zenihiro:2010aa} and even tin isotopes $^{116-124}$Sn \cite{Terashima:2008aa} have been extracted. At high energies $>$500MeV, the analysis is complicated by the production of mesons. However, neutron skin measurements have been extracted from analysis of proton scattering at 650 MeV for $^{208}$Pb \cite{Starodubsky:1994aa} and 800 MeV for $^{48}$Ca, $^{116}$Sn, $^{124}$Sn and $^{208}$Pb \cite{Ray:1979aa}.

At lower energies, the nucleon-nucleon interaction plays a much stronger role and the interpretation of the data more complicated. An analysis of low energy neutron and proton elastic scattering data using ab-initio calculations of $^{48}$Ca has also been carried out \cite{Mahzoon:2017aa}.

\subsubsection{Low energy collective motion}

The oscillations of neutrons in the surface of neutron rich isotopes against the isospin symmetric core (with the symmetry energy slope $L$ - proportional to the pressure of PNM - behaving as a restoring force) are sensitive neutron skin and symmetry energy \cite{Tamii:2011aa,Roca-Maza:2015aa,Birkhan:2017aa}. One way to probe this collective motion is to use 200-400 MeV beams of protons, whose scattering is examined at very forward angles (small momentum transfer). Their angular distribution and polarization transfer amplitudes disentangle the spin-flip M1 and non-spin flip E1 transitions and allow the E1 dipole polarizability to be extracted. Neutron skins of $^{48}$Ca and $^{208}$Pb have been extracted this way.

Using electromagnetic probes, the neutron skins of $^{130,132}$Sn and $^{208}$Pb have been determined by measuring the strengths of the Pygmy Dipole resonances at 5-9MeV \cite{Klimkiewicz:2007aa,Klimkiewicz:2008aa,Carbone:2010aa}, although there is significant uncertainty remaining about the model systematic errors in such analyses \cite{Daoutidis:2011aa}.

\subsubsection{Giant resonances}

At the opposite end to the collective motion energy spectrum, measurement of the cross section of the isovector giant dipole resonance (IVGDR) by alpha scattering \cite{Krasznahorkay:1994aa} in $^{116,124}$Sn and $^{208}$Pb has been used to extract the neutron skin. The neutron skin of $^{208}$Pb has also been inferred from the $\gamma$-decay of the anti-analog of the giant dipole resonance  \cite{Krasznahorkay:2012aa}. The energy of the spin-dipole resonance is sensitive to the neutron skin and has been used to measure the neutron skins of the even tin isotopes $^{114-124}$Sn  \cite{Krasznahorkay:1999aa}

\subsubsection{Coherent pion photoproduction}

Coherent pion production from an electromagnetic beam is sensitive to the total density distribution of the nucleus, and given the accurate knowledge of the proton density distribution can be used to extract the neutron skin. This technique has been applied to $^{48}$Ca \cite{Tarbert:2014aa}. Although this method has the advantage of the initial state being well understood, there is some dispute about the level to which systematic errors in the final state modeling have been underestimated \cite{Miller:2019aa,Gardestig:2015aa}. To represent that here we use errors three times larger than those originally reported \cite{Gardestig:2015aa}.

\begin{figure*}[!t]
\includegraphics[scale=0.37]{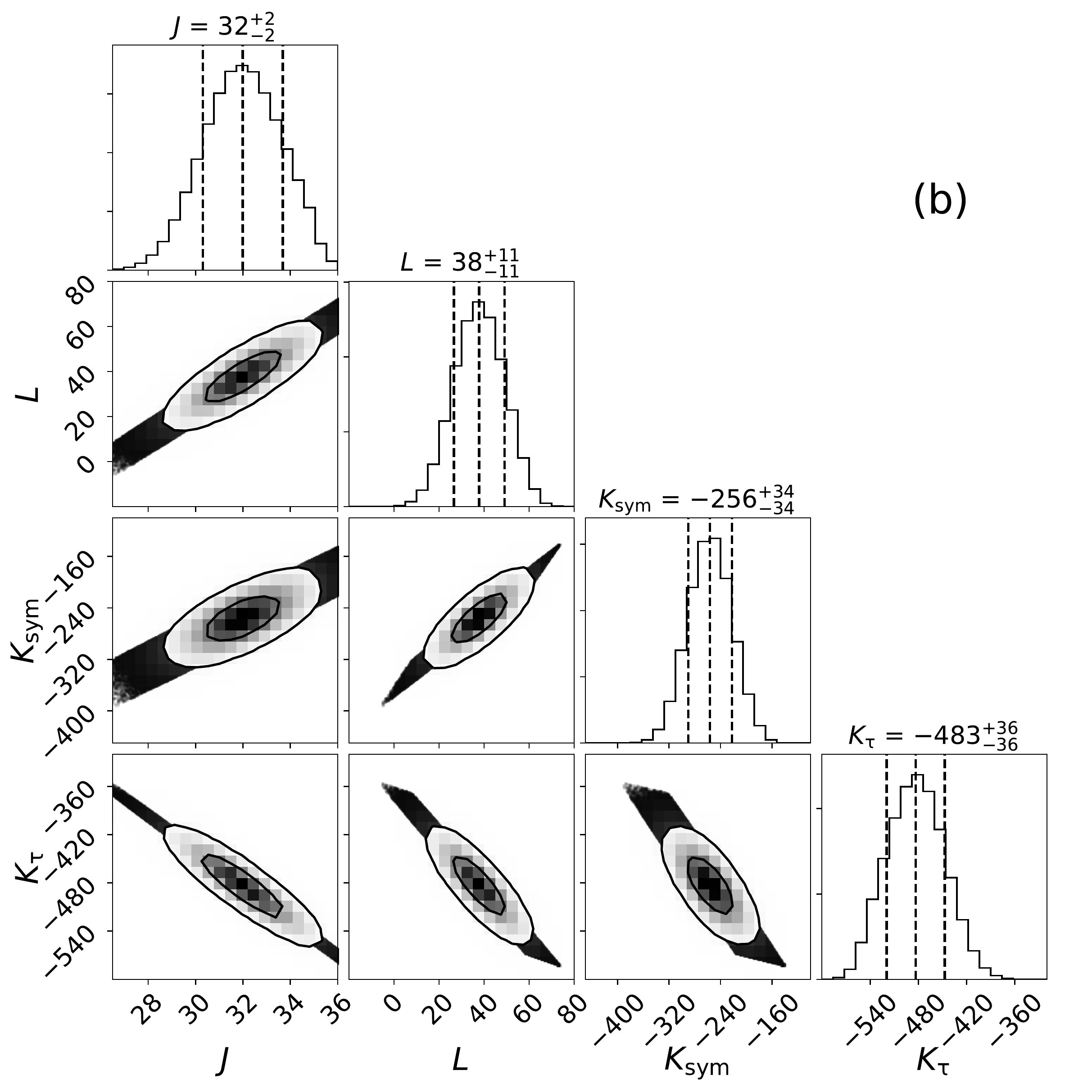}\includegraphics[scale=0.37]{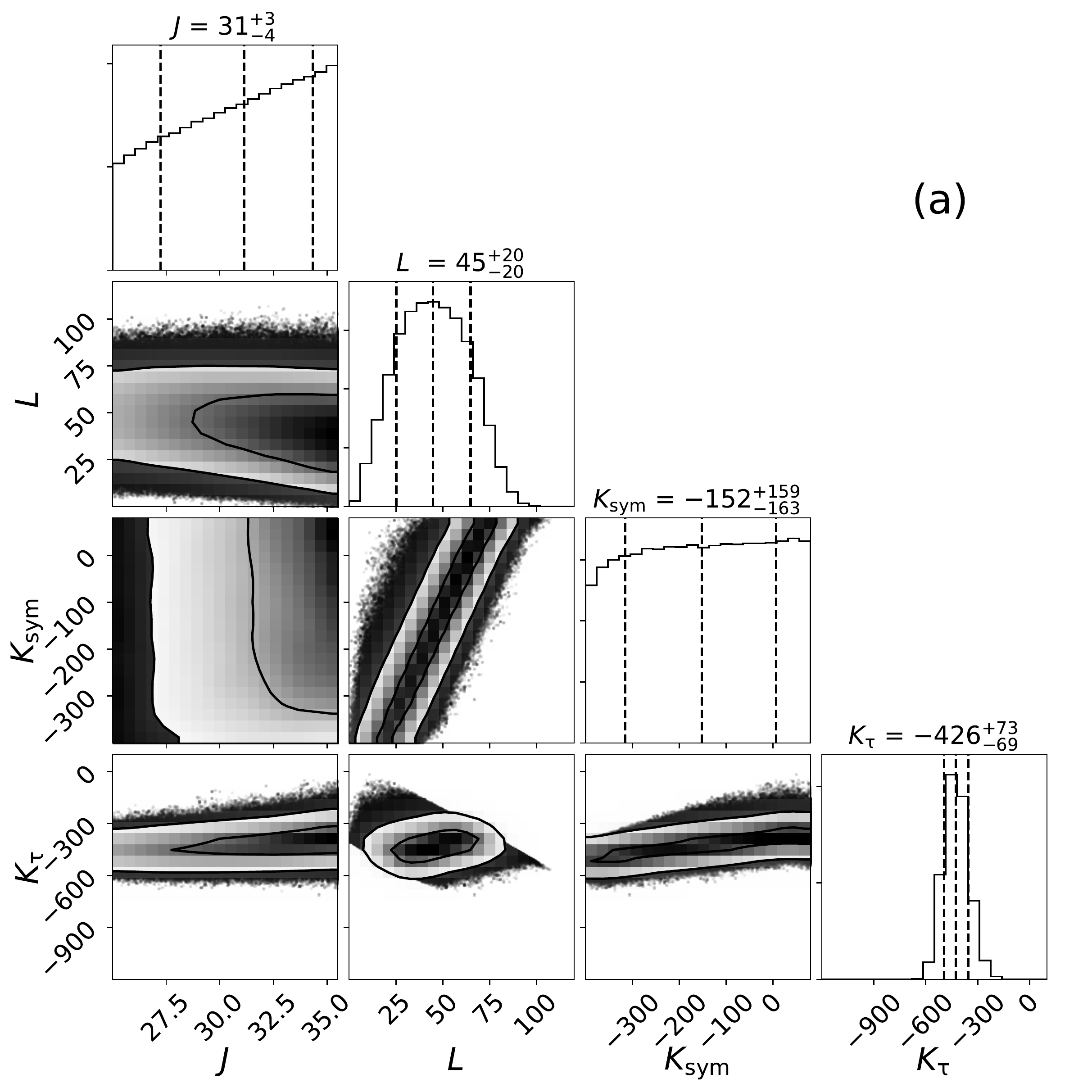}\\
\includegraphics[scale=0.37]{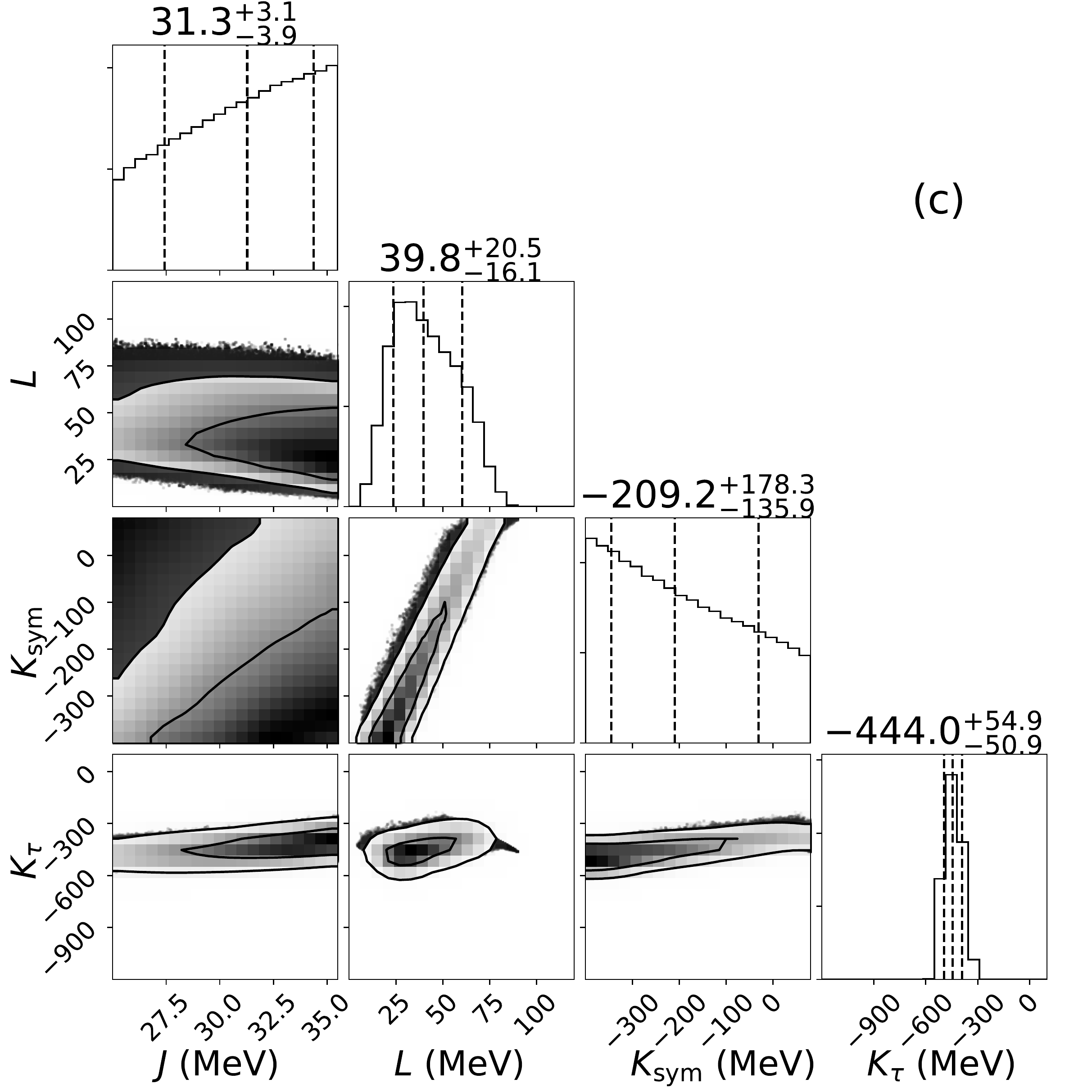}\includegraphics[scale=0.37]{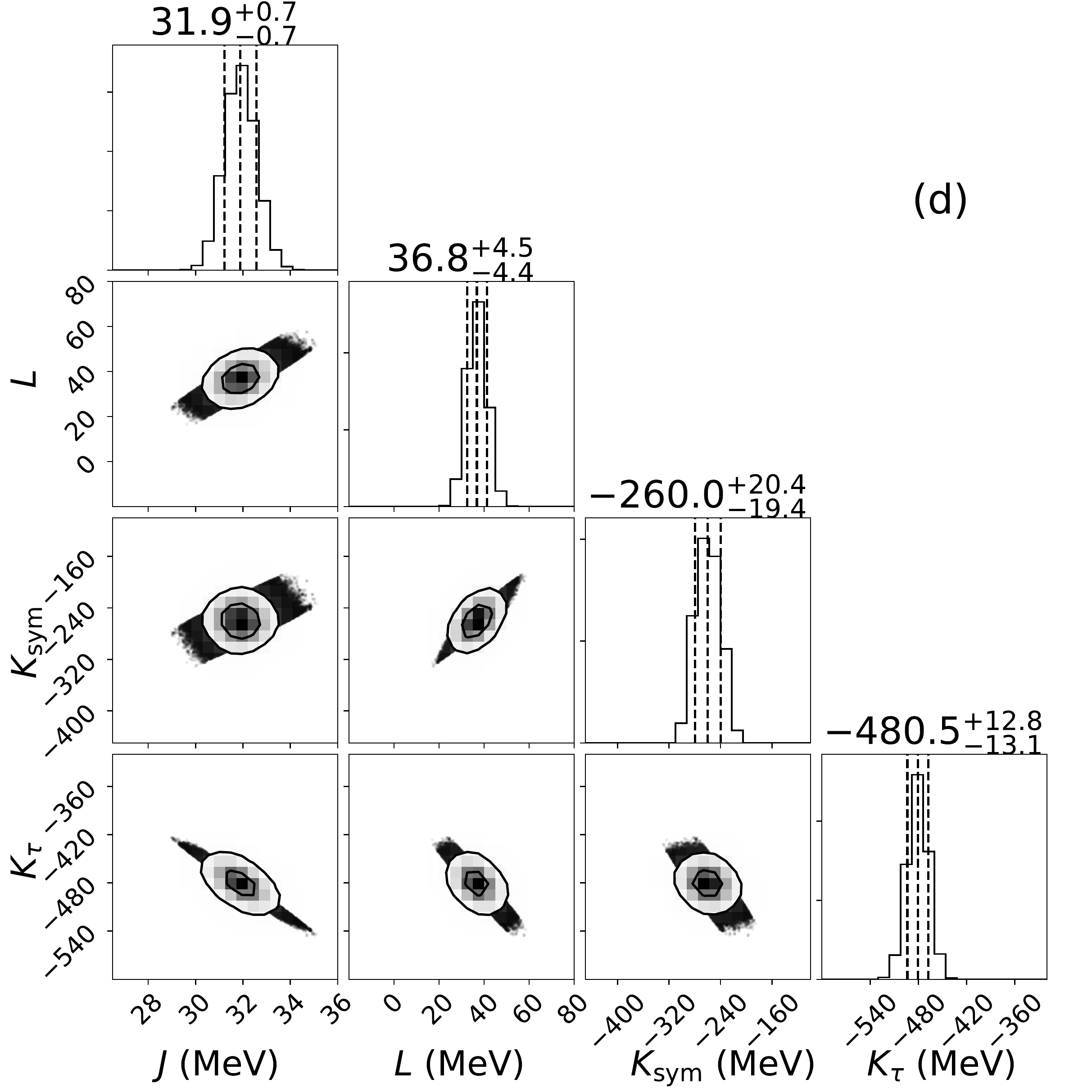}\
\caption{Two-dimensional and one-dimensional marginalized posterior distributions of the symmetry energy parameters $J$, $L$ and $K_{\rm sym}$ as well as the symmetry compressibility $K_{\tau}$, in units of MeV. We show results using uninformative priors (a,c), and pure neutron matter priors (b,d). The top plots (a,b) are the distributions resulting from taking the full range of the error bars on neutron skin data, and the bottom plots are the distributions resulting from adding the errors from different physical probes of the neutron skins in quadrature (c,d). The two contours are the 67\% and 95\% credible regions for the parameters. The reported values at the top of the one-dimensional marginalized distribution are the 67\% credible intervals.} \label{Fig:5}
\end{figure*}

\begin{figure*}[!t]
\includegraphics[scale=0.52]{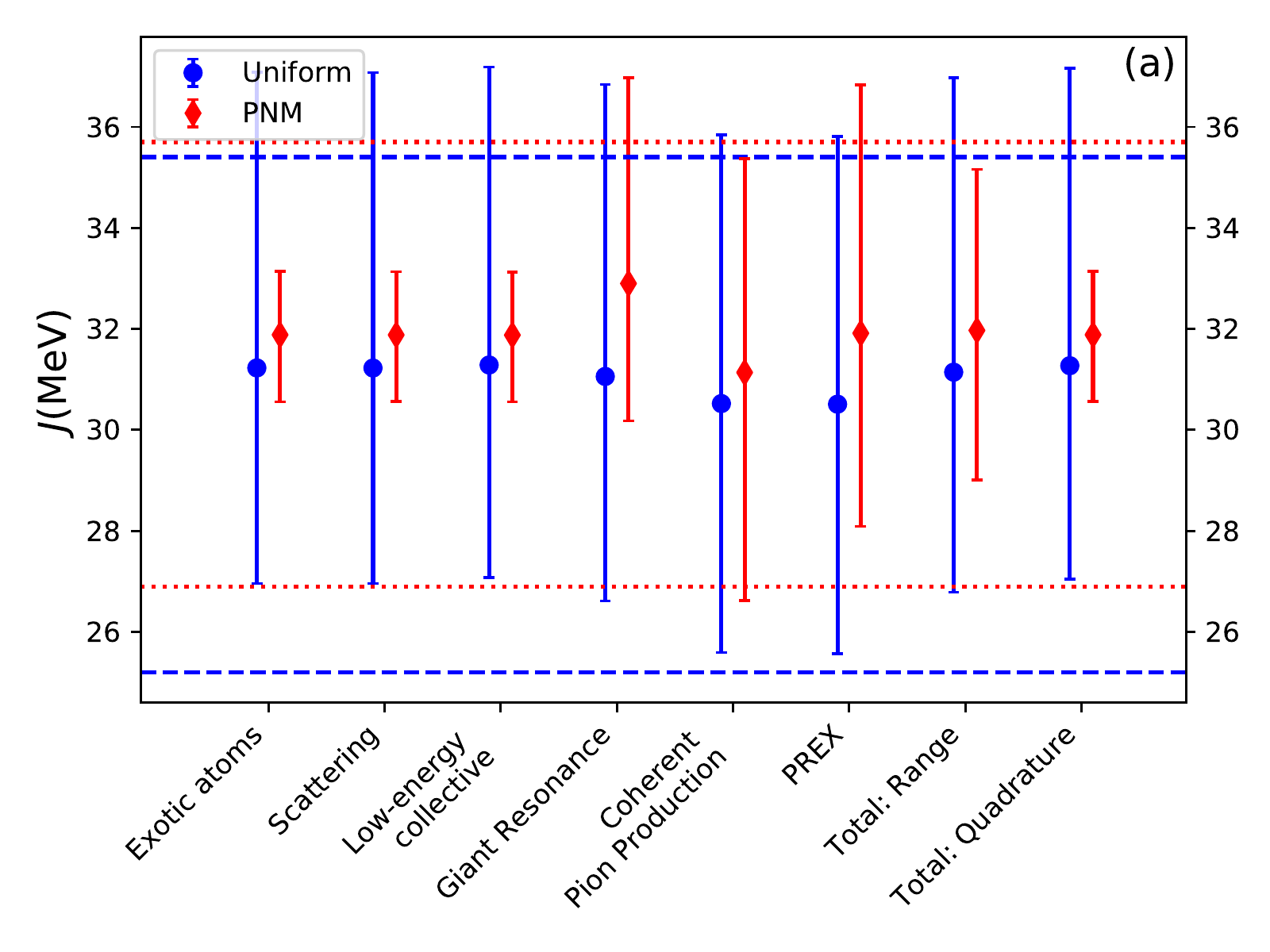}\includegraphics[scale=0.52]{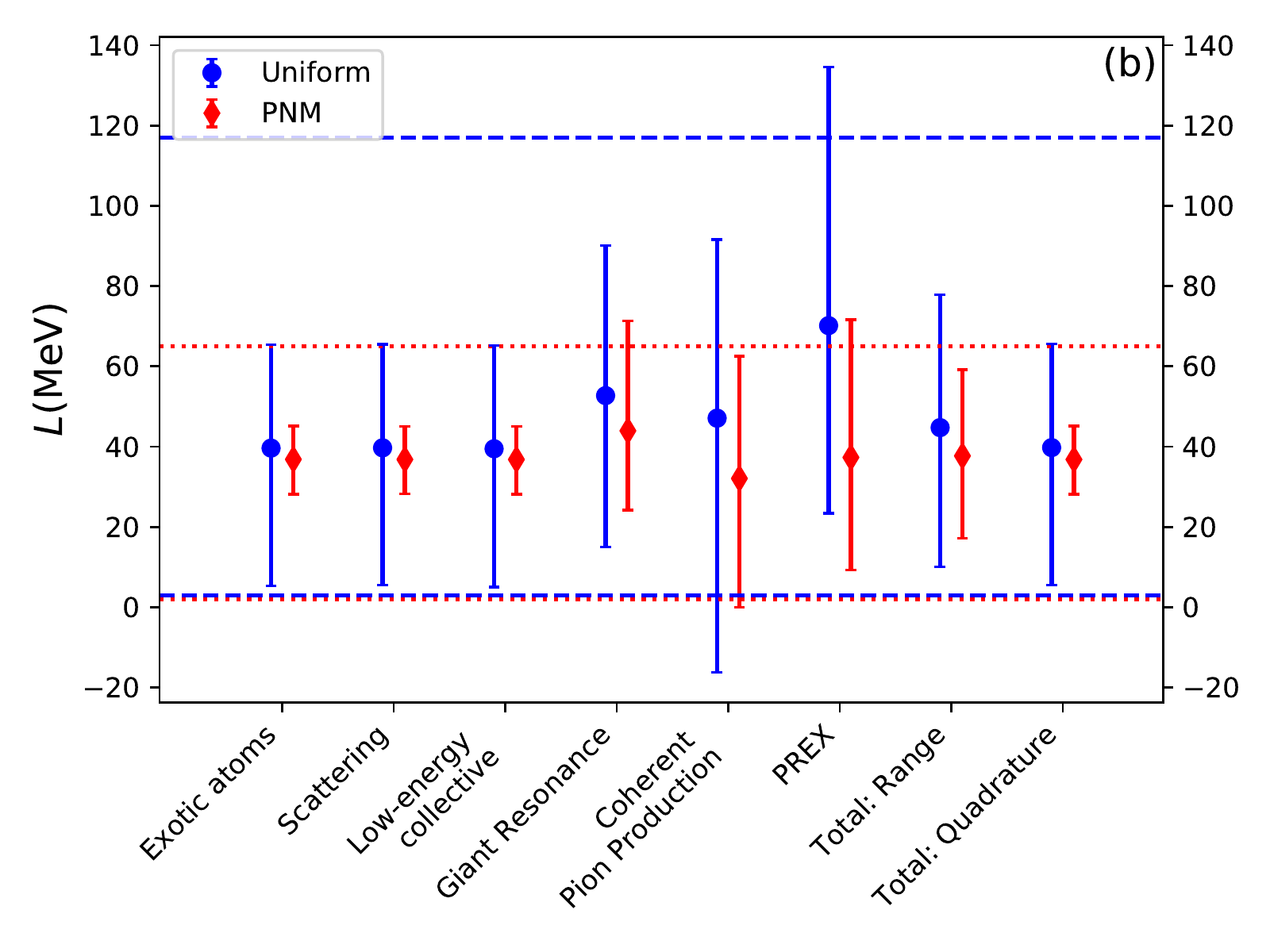}\\
\includegraphics[scale=0.52]{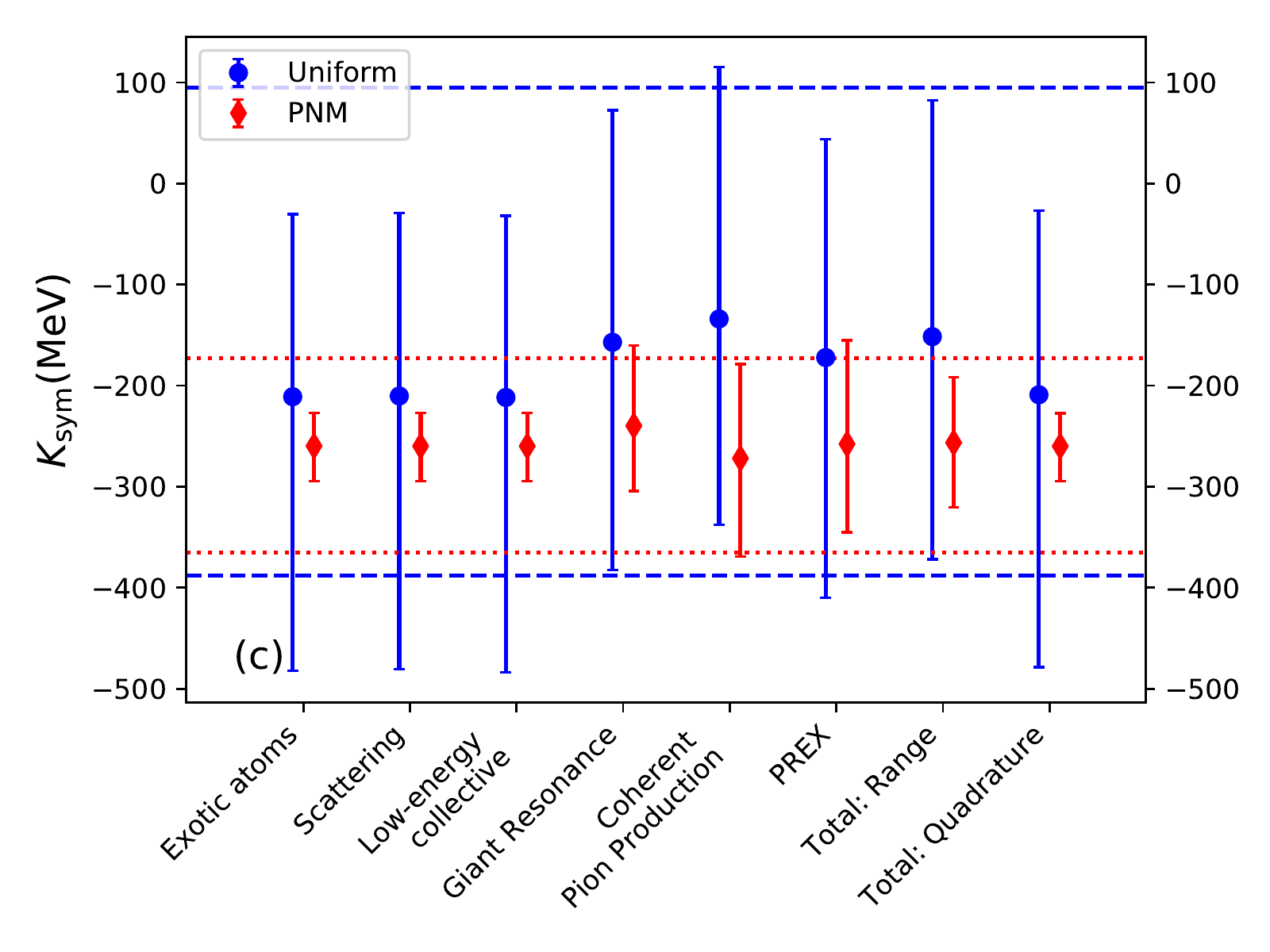}\includegraphics[scale=0.52]{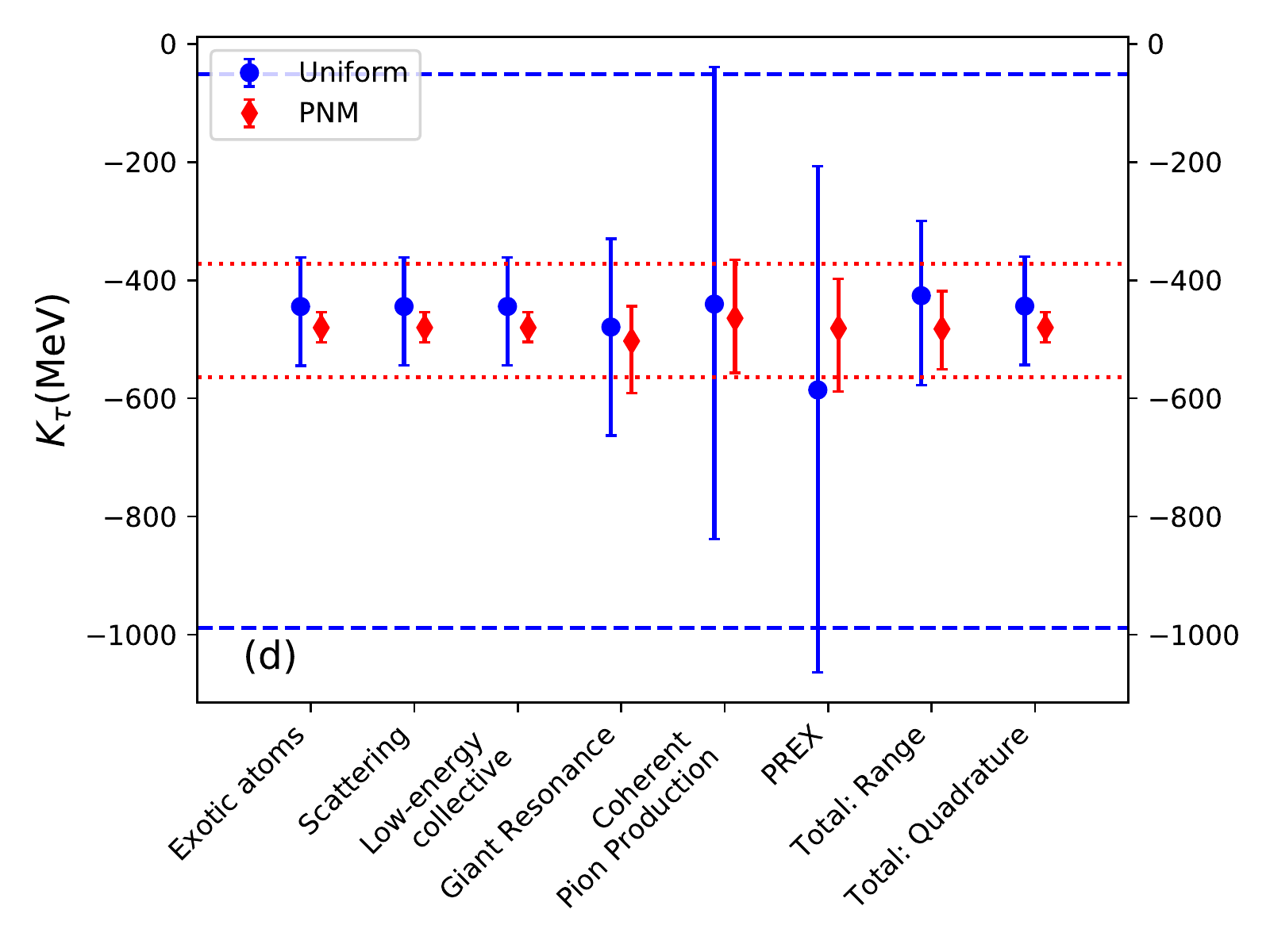}\\
\caption{The 95\% credible error bars on the posterior distributions of the symmetry energy parameters $J$ (a), $L$ (b), $K_{\rm sym}$ (c) and $K_{\tau}$ (d) using the datasets for each individual physical probe and combined as a range and in quadrature. The blue error bars are from our uninformative priors and our red error bars from the pure neutron matter priors. The dashed lines indicate the prior 95\% ranges on our parameters.} \label{Fig:6}
\end{figure*}

\subsubsection{Parity violating electron scattering (PREX)}

Extracting the electroweak signal from electron scattering onto nuclei in principle allows a measurement of the neutron skin independent of the complications of the strong force \cite{Roca-Maza:2011aa,Horowitz:2014aa}. This was carried out once on $^{208}$Pb and although the method proved successful, the statistical significance aimed for was not achieved \cite{Abrahamyan:2012aa}. In the near future, further results on two nuclei, $^{48}$Ca and $^{208}$Pb, will be forthcoming \cite{Becker:2018aa,Thiel:2019aa} \footnote{The PREX-II result was published during revision of this manuscript; it was found to be $0.29 \pm 0.08$fm, which is close to the hypothetical ``stiff'' value we examine in Section VI. Analysis of the PREX-II result will be part of a forthcoming study.}.

\subsubsection{Discussion of data used}

In Figure~4 we plot the data points we use in each of the above categories. Where isotopes have multiple reported measurements in the same category, we combine errors in quadrature.  We compare the data with the ranges obtained from our two sets of priors. The data errors taken are 1$\sigma$ errors, so the ranges for our priors are truncated at the 16th and 84th percentiles to allow us to compare with consistency. The many ways of probing the nuclides and extracting the neutron skins return broadly consistent results which is encouraging; the only place an obvious inconsistency appears is between the proton scattering data of $^{48}$Ca and all other probes.

Given the consistency of these results, it is statistically reasonable to extract single measurements for each nuclide  by adding the results from each category in quadrature, rather than selecting the highest and lowest values across all measurements. In this work, we will compare both ways of obtaining datapoints. However, before we proceed it is worth taking a moment to discuss how we aggregate the more than four decades worth of data we have on neutron skins to assess our current best state of knowledge.

Taking the full range artificially inflates the importance of data from nuclides with relatively few data points; for example, the isotopes $^{112}$Sn, $^{130}$Sn and $^{132}$Sn have only one neutron skin measurement, and have errors of around 0.03 fm. The isotope $^{132}$Sn, with 4 different techniques applied to measuring its neutron skin, has an error range of 0.11 fm, almost four times bigger, because of the greater number of measurements, and certain techniques resulting in larger errors. Lead suffers in this analysis of the data by virtue of having the most attention paid to it: the combination of the large errors of the first PREX measurement and the discussion about the uncertainties in extracting the neutron skin from coherent pion production leads to a total error range of 0.5 fm, which admits well over 90\% of models investigated here. 

A better way to aggregate errors that are reasonably consistent with each other is to add them in quadrature. This rewards those nuclides with the most data, and does not penalize nuclides subject to newer techniques that have larger errors.

Of course, the final, and by far the best way to aggregate the errors is to go back to the actual observables extracted for each experiment, and combine each one in a Bayesian approach using appropriate theoretical modeling to extract the neutron skins from each data set (for example, modeling weak form factors for PREX, optical potentials, and resonance energies and widths with the same set of underlying models). This is well beyond the scope of the current paper, but we would argue is a project it is worth the community to engage with.

\protect
\section{Results}

In Fig~5 we plot the posterior distributions of the symmetry energy parameters $J$, $L$, $K_{\rm sym}$, $K_{\tau}$ for uninformative (a,c) and pure neutron matter priors (b,d). The upper two plots take the data errors to be the highest and lowest values out of all the reported errors and the bottom two plots take the data errors by adding the individual data set errors in quadrature.

\subsection{Uninformative priors}

There are several common features of the posterior distributions. The posterior distribution in $J$ is not strongly peaked, but rather the data prefers higher values of $J$. Similarly $K_{\rm sym}$ is not peaked; the less constraining combined data still gives a roughly uniform posterior while the more constraining combined data prefers smaller values of $K_{\rm sym}$. The slope of the symmetry energy $L$ is most constrained, being peaked around 40-45 MeV in both cases with a width of 20MeV. 

$J$ and $L$ show little correlation. It has been noted that model fits to neutron skin data are expected to induce a positive correlation between $J$ and $L$; we will discuss the reasons why this is not the case in the next section.

There is a positive correlation of $L$ with $K_{\rm sym}$. Therefore, even though $K_{\rm sym}$ is not particularly constrained by the data, the symmetry compressibility $K_{\tau}$ is constrained significantly compared to its prior distribution.

\subsection{PNM priors}

Correlations between $J$ and $L$,$K_{\rm sym}$ inferred by our best knowledge of the PNM EOS lead to very different results. Since a strong correlation between $J$ and $L$ exists for models consistent with microscopic PNM calculations, the neutron skin-induced constraint on $L$ leads to a constraint on $J$ that was not present in the posterior distribution using the uninformative priors. Both $L$ and $K_{\rm sym}$ become significantly more constrained, as does $K_{\tau}$.

One can conclude that the main effect of the neutron skin data is to constrain $L$, induce a correlation between $L$ and $K_{\rm sym}$, and therefore constrain $K_{\tau}$. If additional model correlations are present to start with, $J$ and $K_{\rm sym}$ may be additionally constrained individually. 

\subsection{Comparison of datasets}

To compare the effect of the various datasets on the inferred symmetry energy parameters we plot the 95\% credible intervals in Figure~6. We also indicate the bounds of the prior distribution at the 2.5 and 97.5\% percentiles using blue dashed (uninformative priors) and red dotted (PNM priors) lines. Firstly, let's look at the inferred parameters from the different data sets. The data are very consistent. It's also worth noting that the inferred parameters from the combining data in quadrature do not result in significant gains over the most constraining individual datasets; they are comparable with the inferred values from exotic atoms, scattering data and low energy collective modes (including dipole polarizability) individually. 
These plots emphasize that $J$ and $K_{\rm sym}$ are not constrained using uninformative priors, but starting with PNM priors the values of $J$ become significantly constrained. As expected, the PNM priors lead to more stringent constraints on all parameters, but compared to the prior distributions, both $L$ and $K_{\tau}$ are significantly constrained starting from uninformative priors. It is interesting to note that the neutron skin data constrains $L$ and $K_{\tau}$ starting from our uninformative priors as much as our pure neutron matter prior constrains $L$ and $K_{\tau}$ with no neutron skin data.

\begin{figure}[!t]
\includegraphics[scale=0.36]{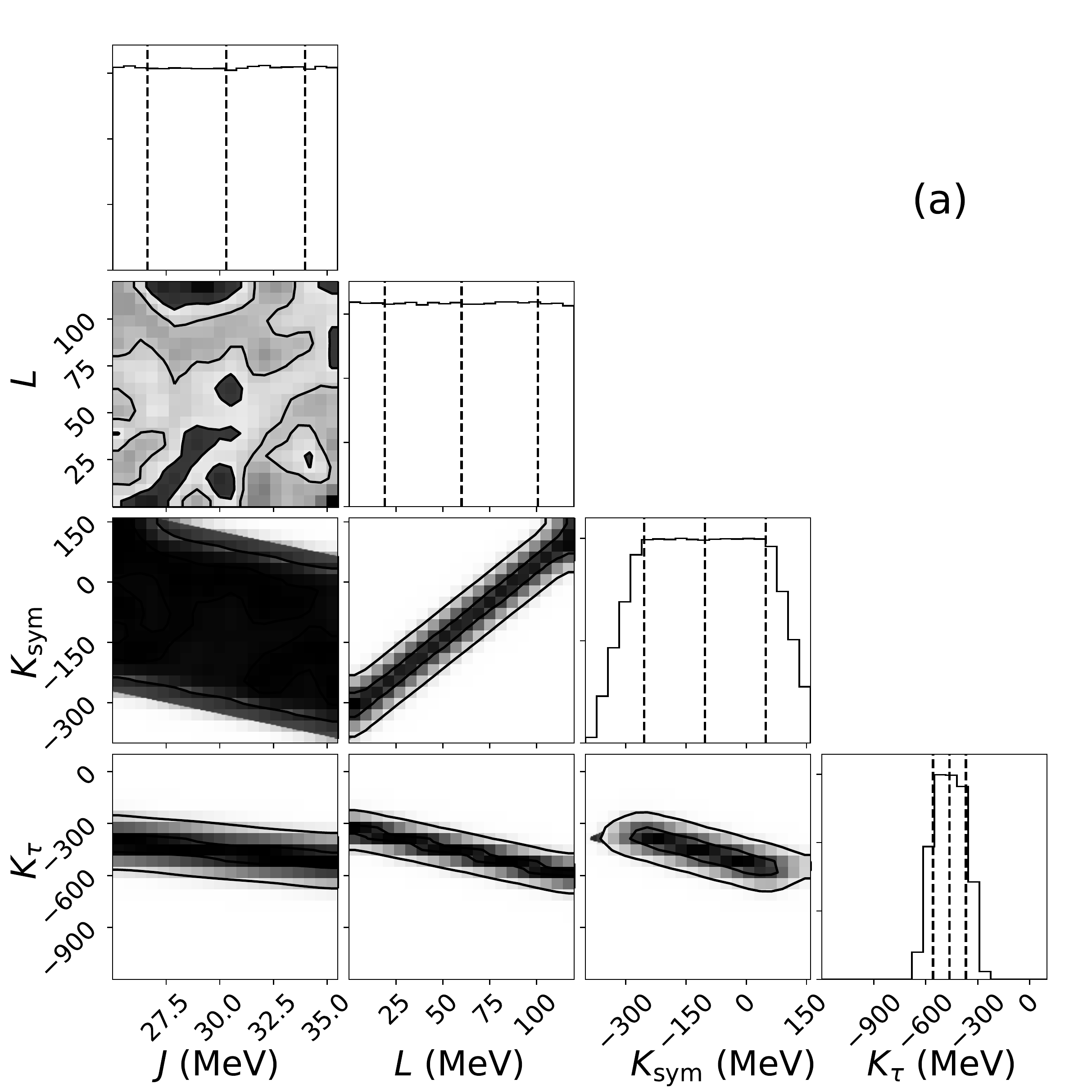}\\
\includegraphics[scale=0.36]{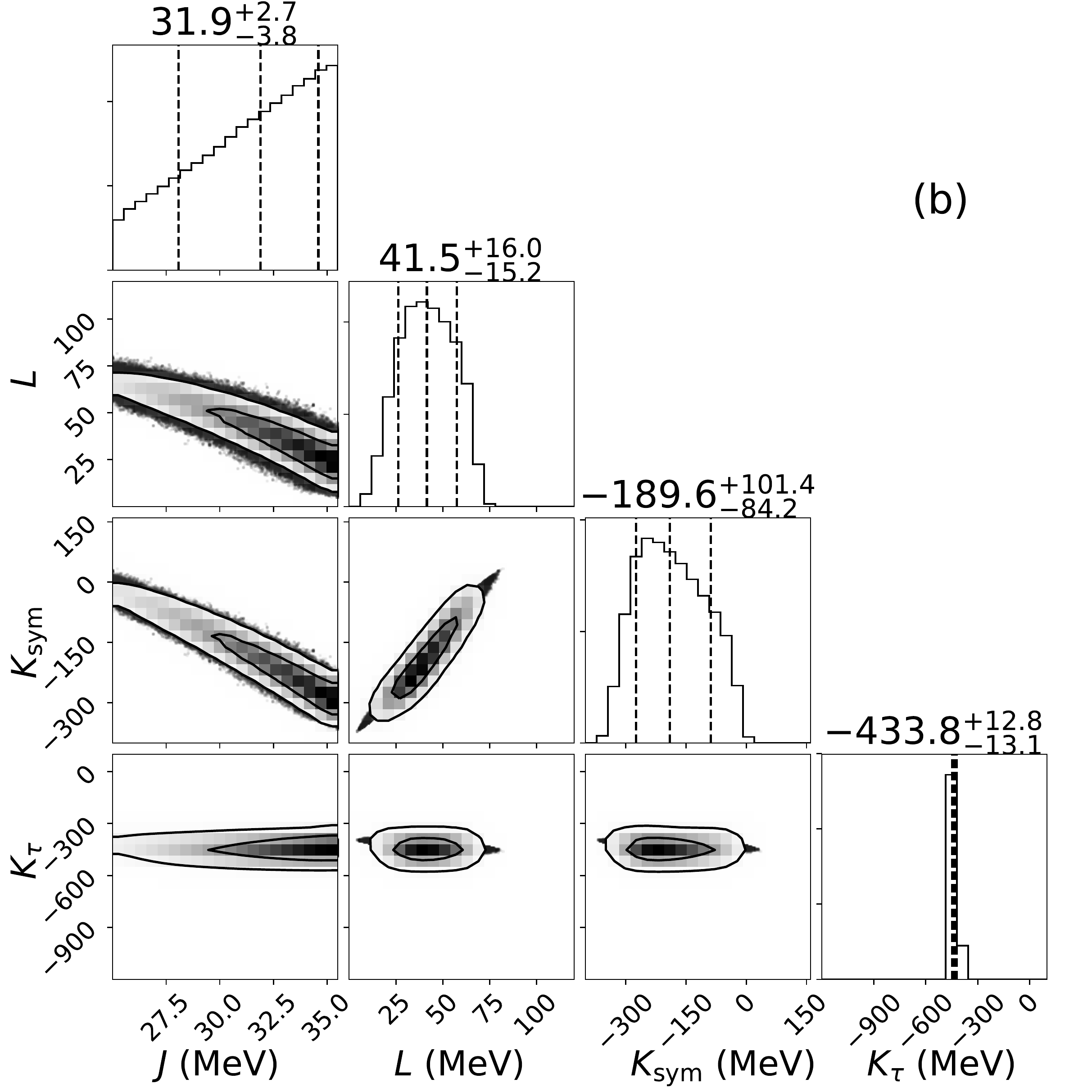}
\caption{Two-dimensional and one-dimensional marginalized posterior distributions of the symmetry energy parameters $J$, $L$ and $K_{\rm sym}$ as well as the symmetry compressibility $K_{\tau}$, in units of MeV, for the MSL0 priors alone (a), and obtained from adding the errors from different physical probes of the neutron skins in quadrature (b). The two contours are the 67\% and 95\% credible regions for the parameters. The reported values at the top of the one-dimensional marginalized distribution are the 67\% credible intervals.} \label{Fig:7}
\end{figure}

\section{Correlations between symmetry energy parameters induced by neutron skin data}

In the analysis of neutron skin data of tin isotopes \cite{Chen:2010aa} the neutron skin data were found to induce a negative correlation between $J$ and $L$. Other authors have noted, based on analysis of the droplet model, that a positive correlation should be induced. Here we seek to resolve this discrepancy.

Firstly, we reproduce the results of \cite{Chen:2010aa}. They used a similar method to ours, except they use a traditional Skyrme in which only $J$ and $L$ were varied independently out of the symmetry energy parameters, and $K_{\rm sym}$ was fixed based on the underlying Skyrme parameter set. In the work of \citet{Chen:2010aa} the MSL0 Skyrme parameterization was used. 

In any traditional Skyrme, $K_{\rm sym}$ would be related linearly to $J$ and $L$ via a relation $K_{\rm sym}=aJ + bL + c$. In the MSL0 interaction, $K_{\rm sym}$ is related to $J$ and $L$ by

\begin{equation}
K_{\rm sym} = 3.71L - 11.13J + 11.93
\end{equation}

We can mimic the results of a traditional Skyrme by introducing another prior: $J$ and $L$ are drawn from a uniform distribution but $K_{\rm sym}$ is determined from equation~20. The corresponding prior distribution, as well as the posterior distributions for the full range of data and the data combined in quadrature are shown in Figure~7. One can see that indeed we reproduce the results of \cite{Chen:2010aa} closely; a negative correlation between $J$ and $L$ is induced by the data. In order to check if this is to be expected, let apply the same data to the droplet model.

\begin{figure*}[!ht]
\includegraphics[scale=0.37]{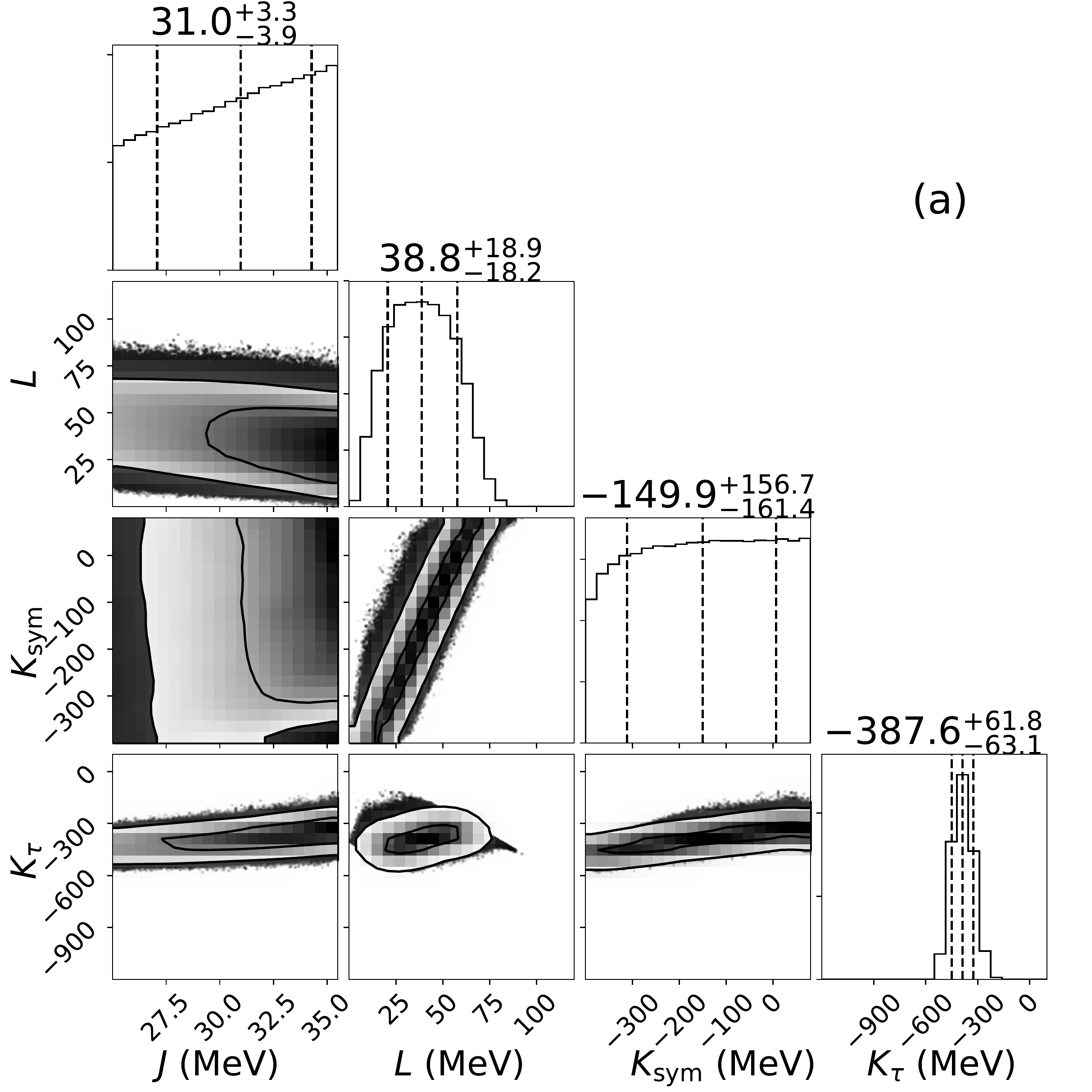}\includegraphics[scale=0.37]{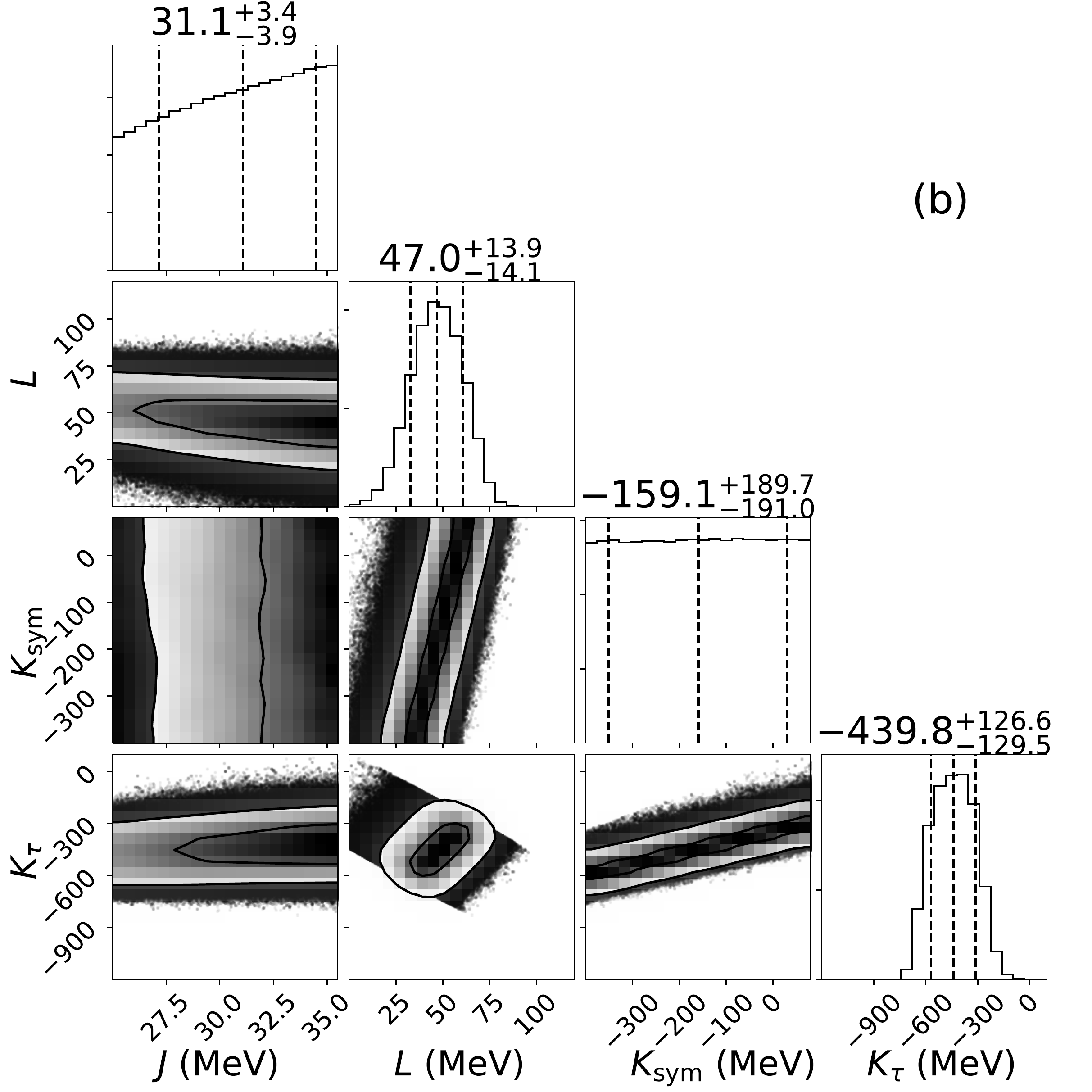}\\
\includegraphics[scale=0.37]{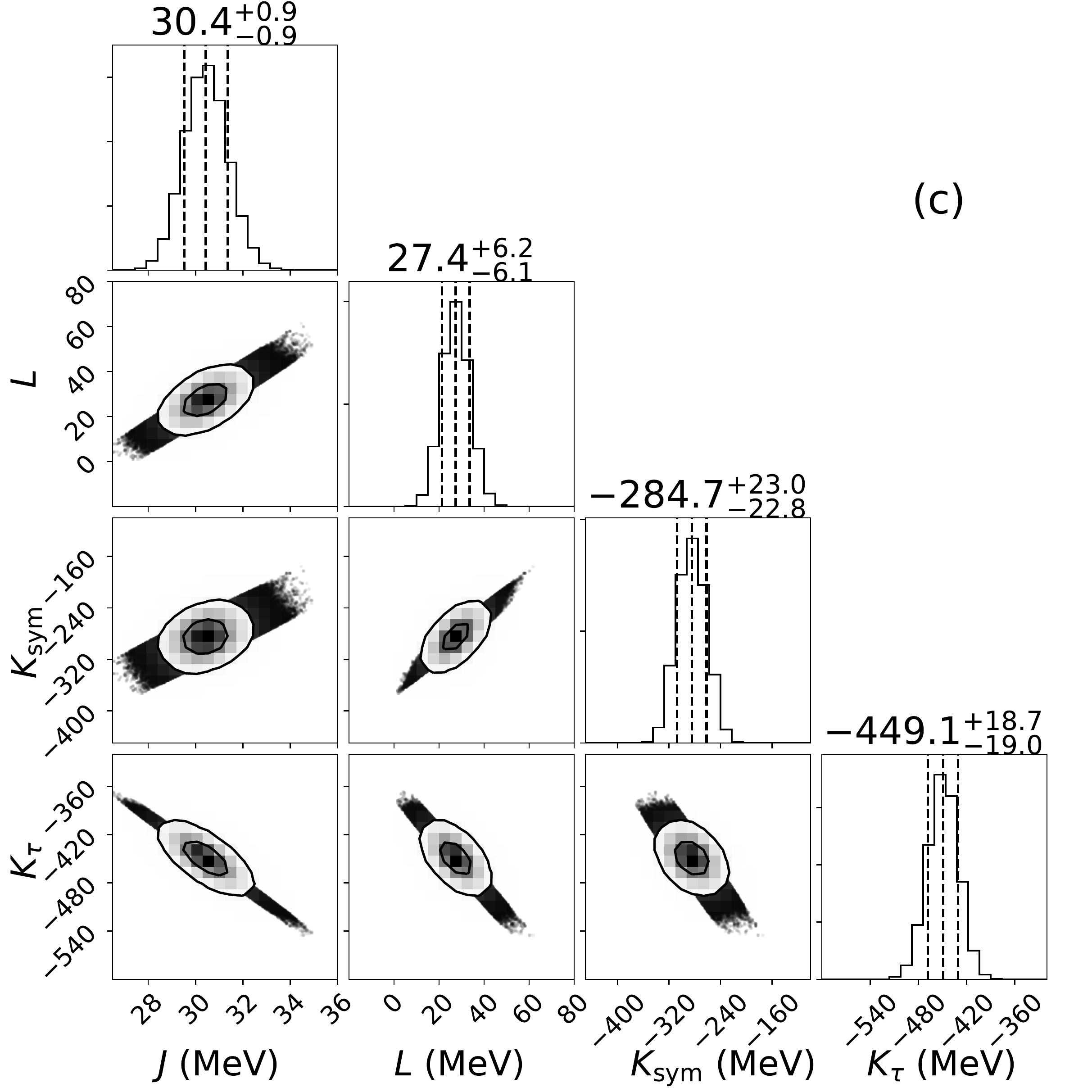}\includegraphics[scale=0.37]{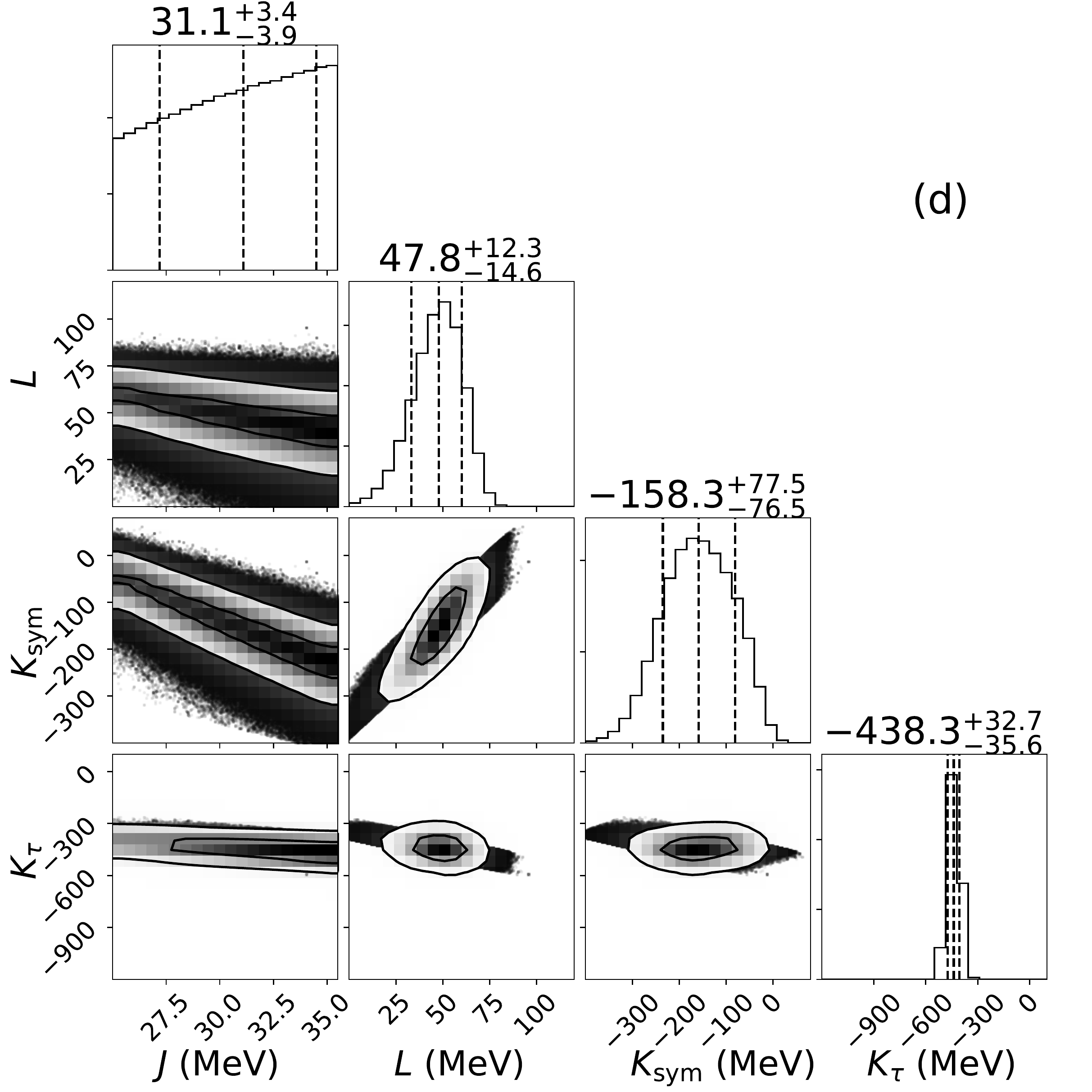}
\caption{Two-dimensional and one-dimensional marginalized posterior distributions of the symmetry energy parameters $J$, $L$ and $K_{\rm sym}$ as well as the symmetry compressibility $K_{\tau}$, in units of MeV, using a hypothetical measurement of the neutron skin of $^{208}$Pb of $0.16 \pm 0.01$ fm. On the left two panels we use the full quantum SHF calculations to model the neutron skins (a,c), whereas on the right two panels we use the droplet model fit equation~(31) (b,d). The top two plots (a,b) are the result of uninformative priors, and the bottom two (c,d) are the result of the MSL0 priors. The droplet model qualitatively reproduces the correlations induced in the symmetry energy parameters of the full SHF model by the neutron skin data. The two contours are the 67\% and 95\% credible regions for the parameters. The reported values at the top of the one-dimensional marginalized distribution are the 67\% credible intervals.} \label{Fig:8}
\end{figure*}

\subsection{The Droplet Model}

It has been noted that based on considerations of the droplet model that neutron skin data ought to induce a positive correlation between $J$ and $L$ \cite{Roca-Maza:2018aa}.

The neutron skin in the droplet model is predicted to have the following form \cite{Myers:1969aa,Brack:1985aa}:

\begin{equation}
    \Delta r_{\rm np} = \sqrt{{3 \over 5}} \bigg( t_{\rm np} - {e^2 Z \over 70 J} \bigg) + \sigma_{\rm sw} \delta
\end{equation}

\noindent where $t_{\rm np}$ is the difference between the mean locations of the neutron and proton surfaces in the presence of just strong interactions, the second term is the Coulomb correction and the third term is a correction due to the difference in the surface thicknesses of the neutron and proton distributions. In the droplet model,

\begin{equation}
    t_{\rm np} = {2 r_0 \over 3 J} [ J - a_{\rm sym}(A)] A^{1/3} (\delta - \delta_{\rm c}),
\end{equation}

\noindent where $r_0$ is the nuclear radius parameter, $\delta = (N-Z)/A$, $\delta_{\rm c}=(e^2 Z / 20 J)A^{-1/3}$ and $a_{\rm sym}(A)$ is the surface symmetry energy:

\begin{equation}
    a_{\rm sym}(A) = J \bigg( 1 + {9J \over 4Q}A^{-1/3} \bigg)^{-1}.
\end{equation}

It has been shown \cite{Centelles:2009aa} that to a good approximation, $a_{\rm sym}(A)$ is equal to the bulk symmetry energy at some sub-saturation density $\rho_{\rm A}$:

\begin{equation}
    a_{\rm sym}(A) = S(\rho_{\rm A}) = J + \chi_{\rm A}L + {\chi_{\rm A}^2 \over 2} K_{\rm sym} + \dots,
\end{equation}

\noindent where $\chi_{\rm A} = (\rho_{\rm A} - \rho_0) / 3 \rho_0$. This allows us to write $Q$ in terms of bulk symmetry energy parameters: defining

\begin{equation}
r_{\rm sym}(\rho_{\rm A}) = J / S(\rho_{\rm A}),
\end{equation}

\noindent we can write

\begin{equation}
    {J \over Q} = {4 \over 9}(r_{\rm sym} - 1) A^{1/3}.
\end{equation}

Then $r_{\rm sym}$ and $\rho_{\rm A}$ are model parameters that contain the information about the density dependence of the symmetry energy. We have

\begin{equation}
        t_{\rm np} = {3 \over 2} A^{1/3} {J \over Q} {\delta - \delta_{\rm c} \over r_{\rm sym}}.
\end{equation}

For the surface thickness contribution, it has been shown \cite{Warda:2009aa} that $\sigma_{\rm sw}$ can be parameterized by

\begin{equation}
    \sigma_{\rm sw} = a {J \over Q} + b,
\end{equation}

\noindent with $a$ and $b$ that have previously been fit only to semi-infinite nuclear matter calculations from a limited number of density functionals \cite{Warda:2009aa}.

Overall, emphasizing the symmetry energy content, the droplet model expression for the neutron skin is

\begin{widetext}
\begin{equation}
    \Delta r_{\rm np} = \sqrt{{3 \over 5}} \bigg[ {3 \over 2} \bigg({J \over Q} {\delta \over r_{\rm sym}} - {J \over Q}  {1 \over J r_{\rm sym}} {e^2 Z \over 20} A^{-1/3}\bigg) - {1 \over J} {e^2 Z \over 70} \bigg] + a {J \over Q} \delta + b \delta.
\end{equation}
\end{widetext}

Based on this, we fit the results of our SHF calculations to the following parameterization of the neutron skin:

\begin{widetext}
\begin{equation}
    \Delta r_{\rm np} = a_1 {J \over Q} {1 \over r_{\rm sym}(\rho_A)}  \delta + a_2 {J\over Q} {1 \over J r_{\rm sym}(\rho_A)} {Z A^{-1/3}}   + a_3 {1 \over J} Z + a_4 {J \over Q} \delta + a_5 \delta.
\end{equation}
\end{widetext}

\noindent Here, $a_{1-5}$ and $\rho_A$ are parameters fit to our SHF calculations and we use the symmetry energy expansion to calculate $S(\rho_{\rm A})$. Performing a non-linear least-squares fit to the results of the Skyrme-Hartree-Fock calculations, we get $\rho_A=0.0960$fm$^{-3}\pm 0.001$, $a_1$=$1.065 \pm 0.11$, $a_2=-0.054 \pm 0.0085$, $a_3=-0.0375 \pm 0.0077$, $a_4=0.362 \pm 0.037$ and $a_5=0.758 \pm 0.101$. 


Comparing with the values calculated directly from expression~(29), $a_1=1.16$, $a_2=-0.0837$, $a_3=-0.0159$, $a_4$=0.3 and $a_5=-0.05$ to $0.07$, we see that the two Coulomb parameters $a_2$ and $a_3$ don't agree particularly well within the errors, and the final term $a_5$ is larger that that found in \cite{Warda:2009aa}. If we hold $a_2$ and $a_3$ fixed at the droplet model values, we get a similar quality of fit, however, and the exact numbers do not affect the analysis to follow. It is for a future work to analyze how well the droplet model can reproduce such an extensive set of quantum calculations of neutron skins; here we want to understand the correlations that are emerging qualitatively.

Using the best fit numbers, we conduct a Bayesian inference of the symmetry energy parameters from the droplet model. We use a hypothetical measurement of the neutron skin of lead of $\Delta r =0.16 \pm 0.01$fm. In Figure~8 we compare the results inferred using the SHF simulations (left two plots) to those from the droplet model fit (right two plots) for uninformative priors (top two plots) and for the MSL0 priors (bottom two plots).

The key takeaway is our droplet model fit qualitatively reproduces the results of the full quantum calculations, \emph{including the negative correlation between $J$ and $L$ induced by the neutron skin data}. We have found this to be true also if one neglects the last two surface thickness terms in equation~(31).

\begin{figure*}[!t]
\includegraphics[scale=0.3]{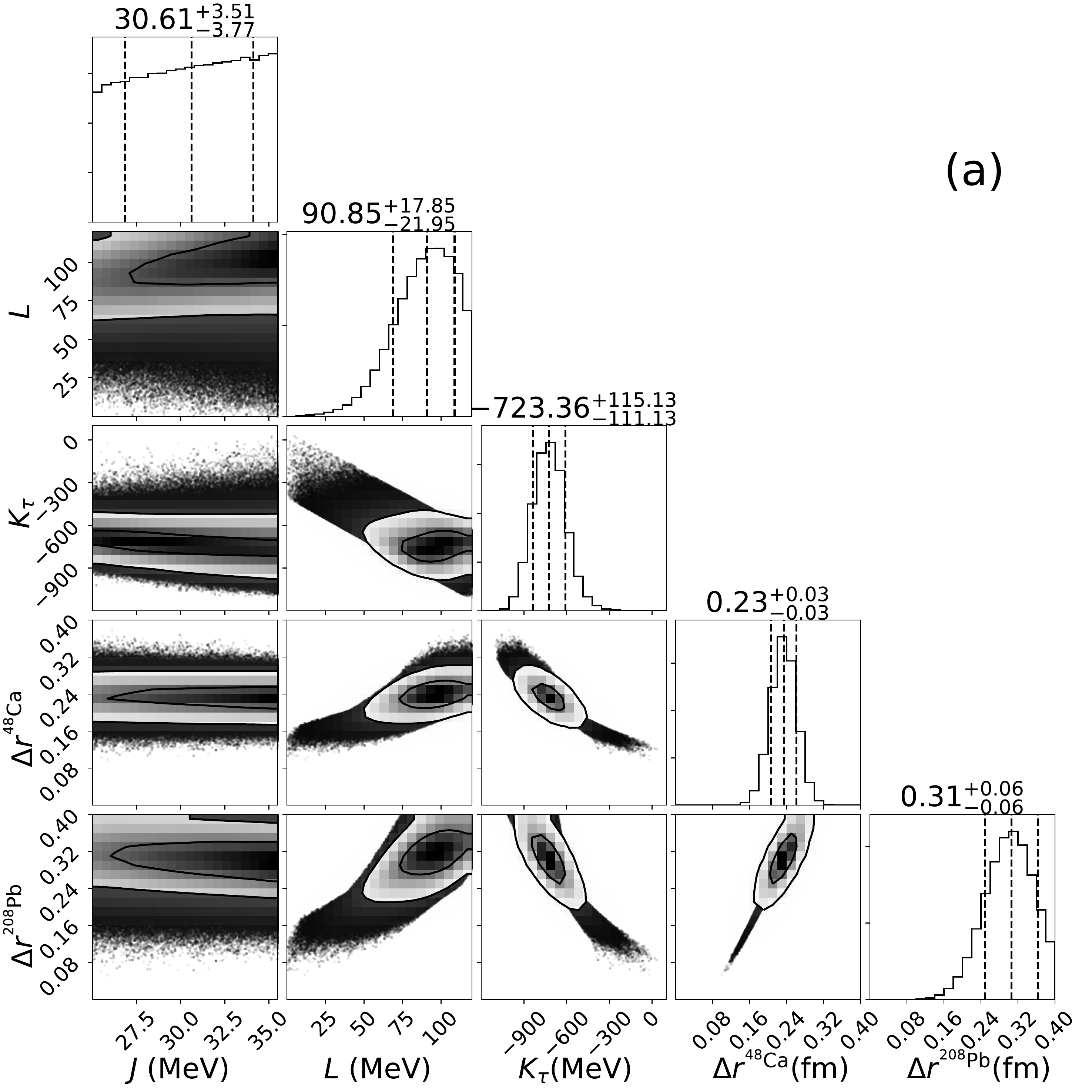}\includegraphics[scale=0.3]{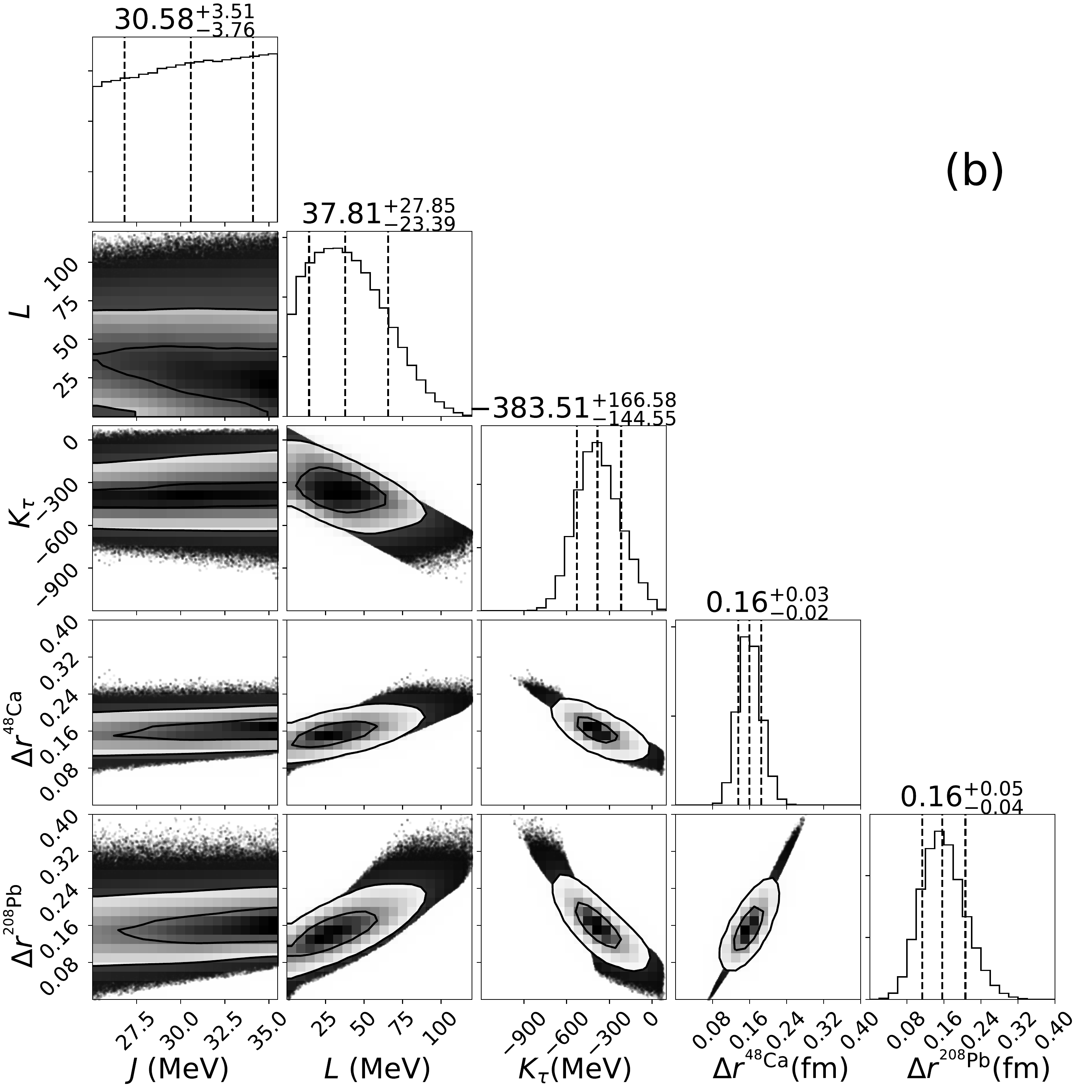}\\
\includegraphics[scale=0.3]{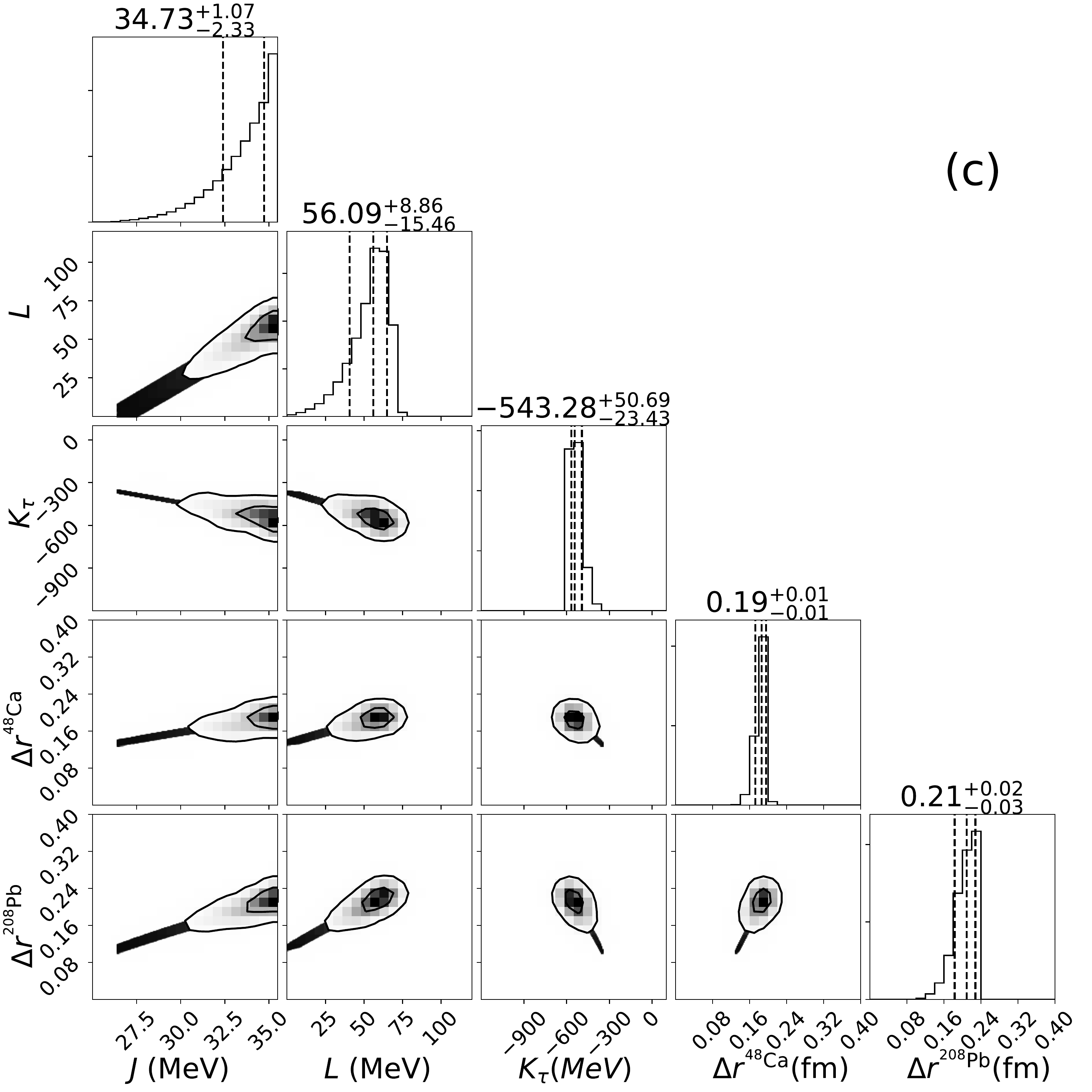}\includegraphics[scale=0.3]{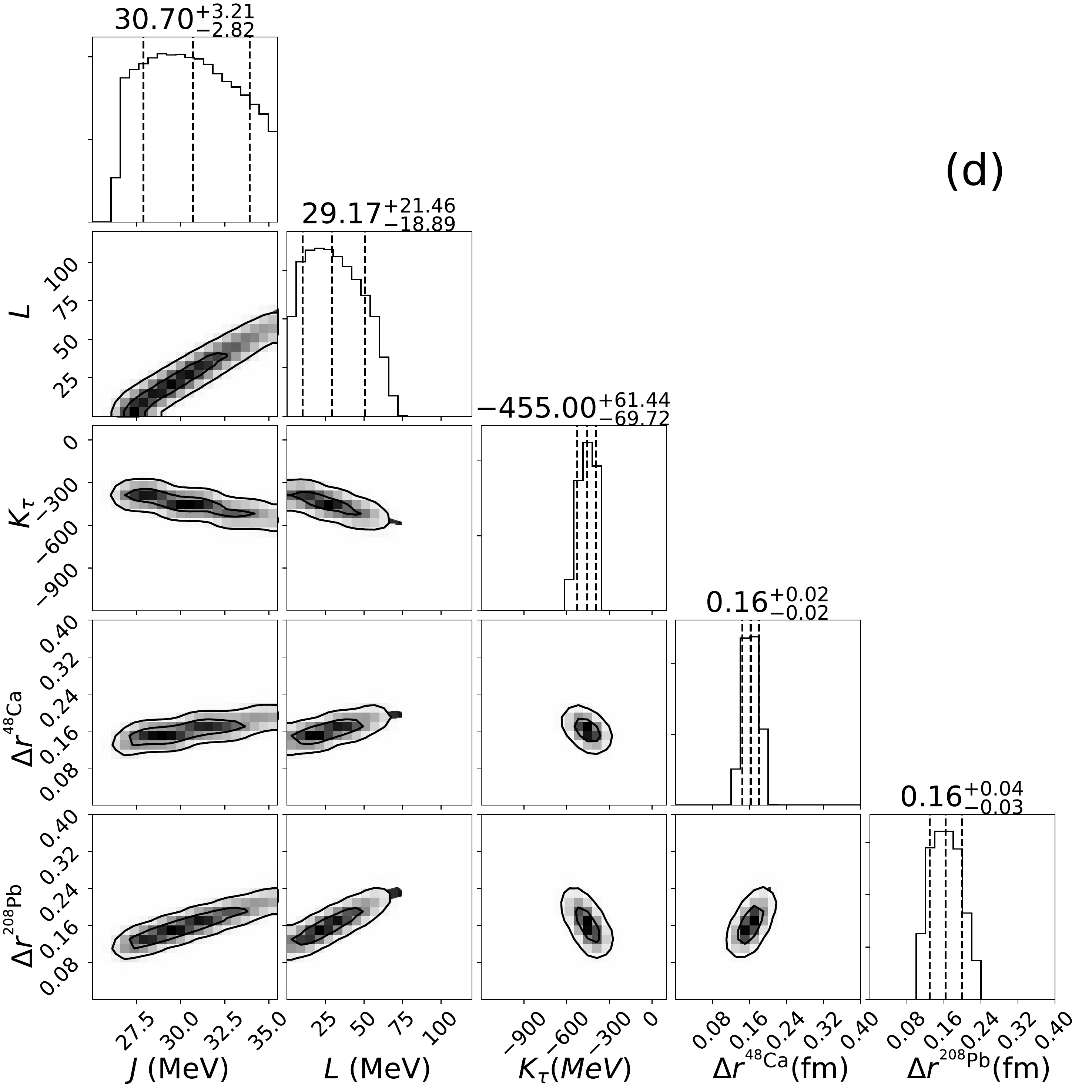}
\caption{Two-dimensional and one-dimensional marginalized posterior distributions of the symmetry energy parameters $J$, $L$ and $K_{\tau}$, in units of MeV, and the neutron skin thicknesses of $^{208}$Pb and $^{48}$Ca, in units of fm, using for two sample measurements of the neutron skin of $^{208}$Pb of $0.33 \pm 0.06$fm (a,c) and $0.15 \pm 0.06$fm (b,d). We show results from our uninformative priors (a,b) and PNM priors (c,d). The two contours are the 67\% and 95\% credible regions for the parameters. The reported values at the top of the one-dimensional marginalized distribution are the 67\% credible intervals.} \label{Fig:9}
\end{figure*}

\section{Future PREX results}

The original goal for the PREX experiment was to measure the parity-violating asymmetry $A_{\rm pv}$ to an accuracy of 3\% \cite{Roca-Maza:2011aa}. This was estimated to lead to an inference of the neutron skin of lead with an error of $\pm 0.06$ fm (based on the models used). A linear correlation between $\Delta r_{\rm np}$ and $L$ then leads to a predicted error on $L$ of approximately $\pm 40$ MeV. In Figure~9 we test what a measurement with an error of $\pm 0.06$ would imply about the symmetry energy distributions of our models. We take two sample measurements, one with a mean at the same value as PREX, 0.33fm, and one with a soft neutron skin of 0.15fm. This time we show the marginalized posterior distributions of $J$, $L$, $K_{\tau}$ and the posterior distributions of $\Delta r_{\rm np}^{^{208}\rm Pb}$ and $\Delta r_{\rm np}^{^{48}\rm Ca}$. For the uninformative priors, the 67\% credible intervals for $L$ and $K_{\tau}$ are 91$\substack{+18 \\ -22}$ MeV and -721$\substack{+114 \\ -110}$ MeV for the ``High'' measurement and -38$\substack{+27 \\ -23}$ MeV and -386$\substack{+167 \\ -146}$ MeV for the ``Low'' measurement. For PNM priors, we have 58$\substack{+10 \\ -15}$ MeV and -550$\substack{+50 \\ -22}$ MeV for the ``High'' measurement and 30$\substack{+26 \\ -22}$ MeV and -456$\substack{+71 \\ -84}$ MeV for the ``Low'' measurement. 

\begin{figure*}[!t]
\includegraphics[scale=0.37]{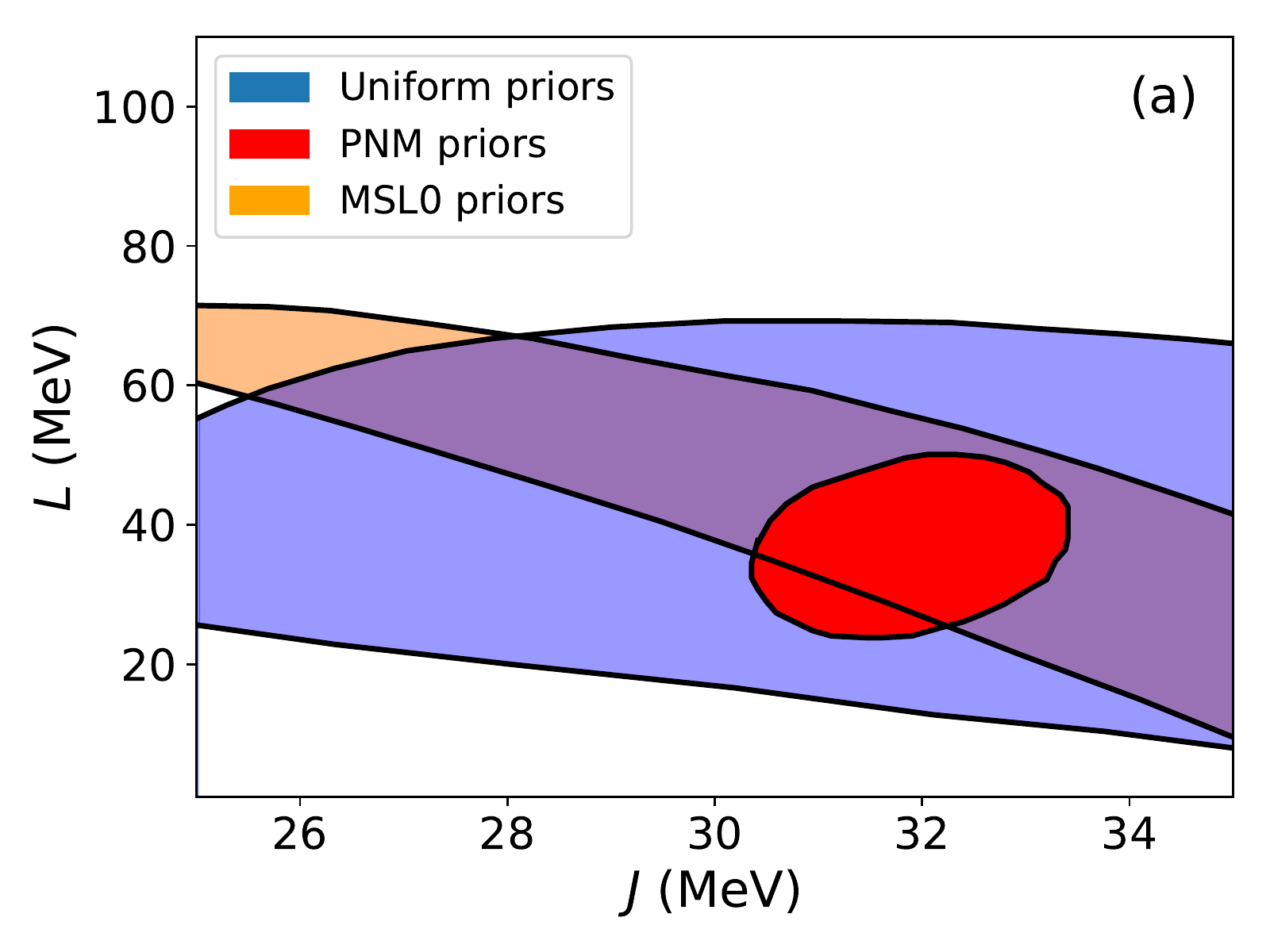}\includegraphics[scale=0.37]{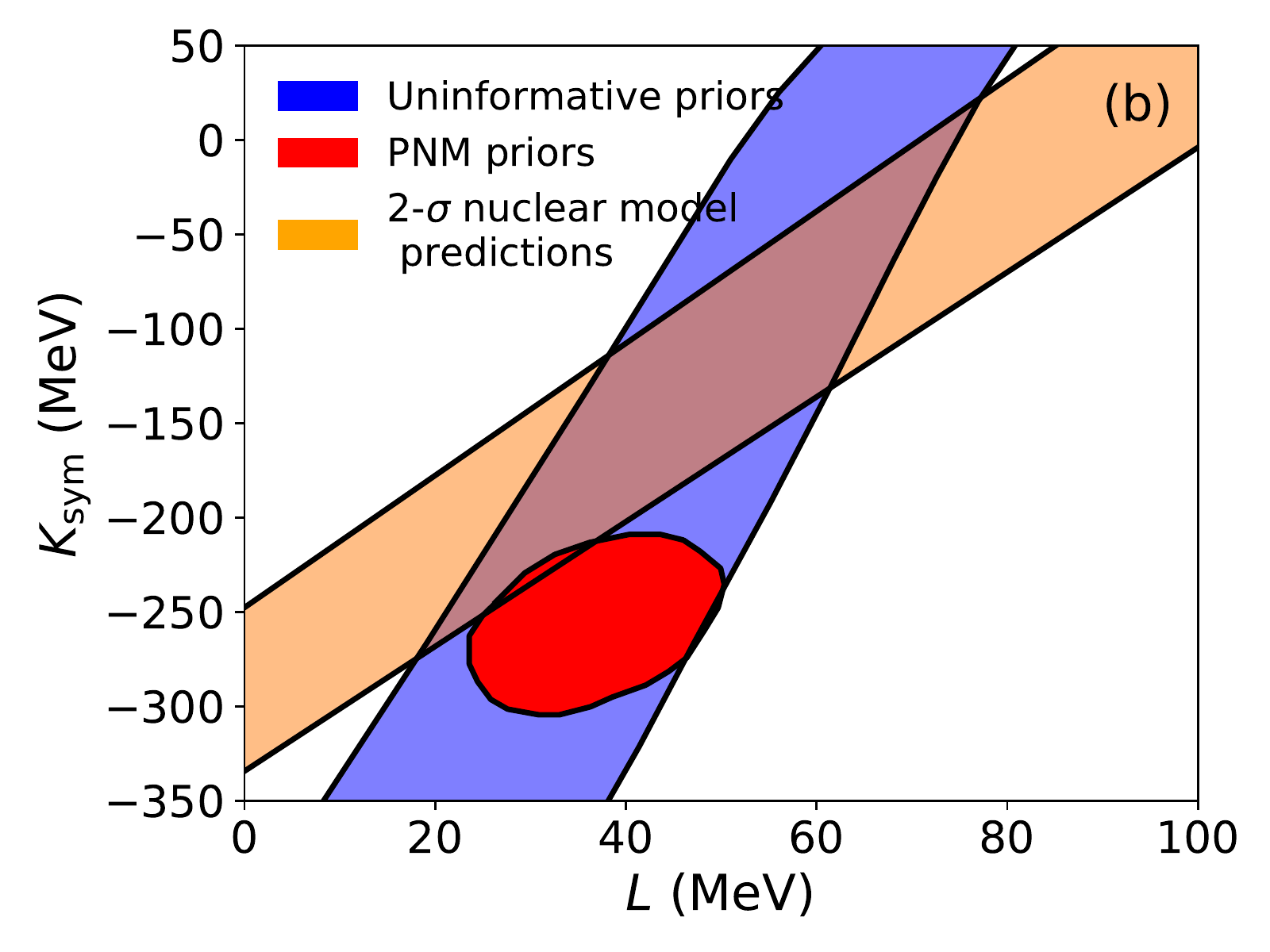}\includegraphics[scale=0.37]{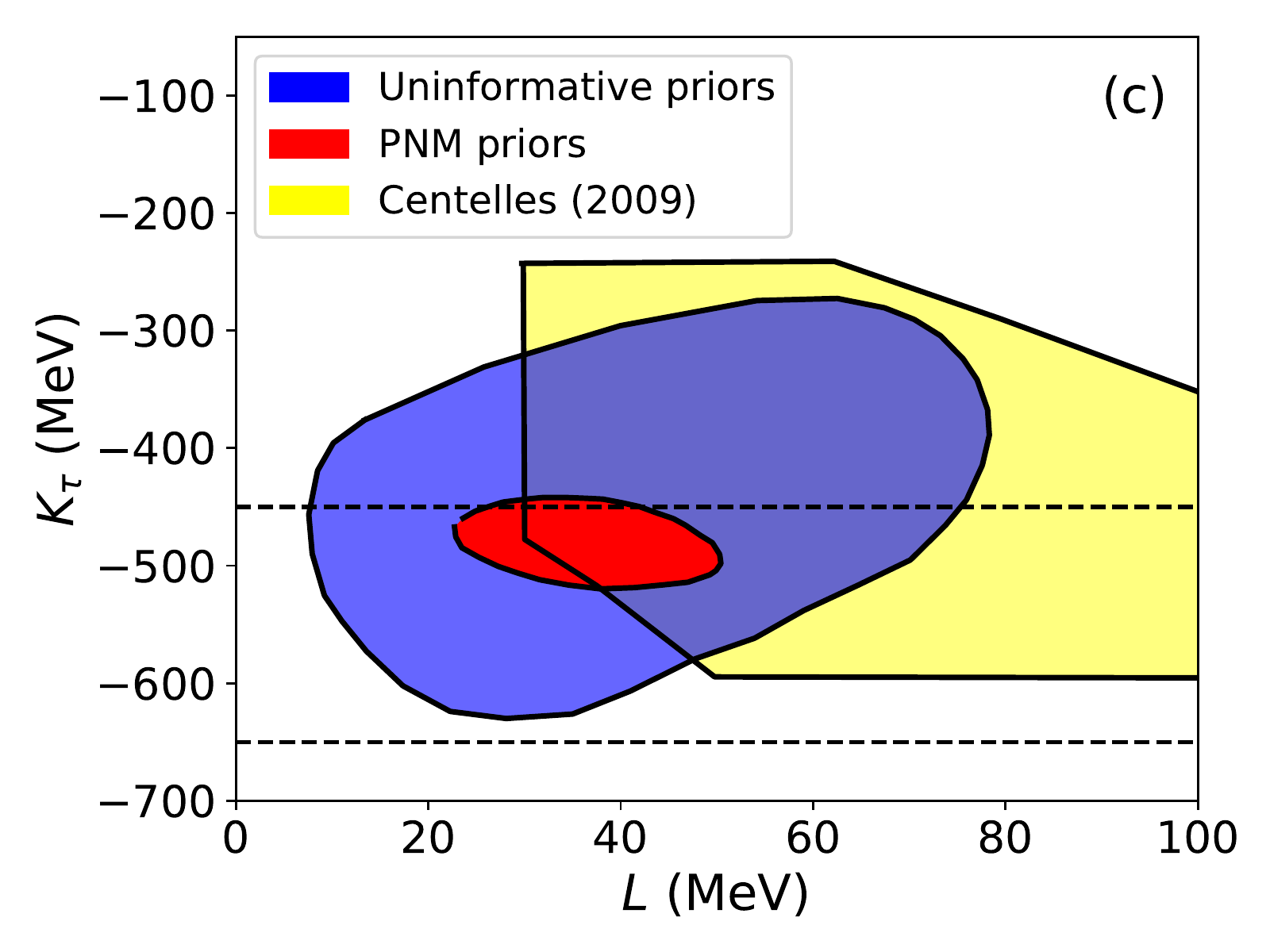}
\caption{Summary of constraints on symmetry energy parameters. In Figure 10a, the 95\% credible regions of the $J$ and $L$ plane are shown for uninformative (blue), PNM (red), and MSL0 (orange) priors. In 10b, the 95\% credible regions of the $L$ and $K_{\rm sym}$ plane are shown for uninformative (blue) and PNM (red) priors; the orange region is bounded by the $2\sigma$ limits of a $\chi^2$ fit to several hundred existing nuclear models \cite{Tews:2017aa}, and appears inconsistent with the neutron skin data analyzed with PNM priors. In 10c, the 95\% credible regions of the $L$ and $K_{\tau}$ plane are shown for uninformative (blue) and PNM (red) priors; the yellow region is the estimated $1\sigma$ bounds from a droplet model analysis \cite{Centelles:2009aa} and the dashed lines bound $1\sigma$ results from the analysis of heavy ion collision and giant resonance data \cite{Chen:2005aa,Shetty:2007aa,Li:2007aa}.} \label{Fig:10}
\end{figure*}

The CREX experiment aims to achieve limits of $\pm 0.02$ fm. Taking a sample measurement of $0.15\pm0.02$, for the uninformative priors, the 67\% credible intervals for $L$ and $K_{\tau}$ are 33$\substack{+24 \\ -19}$ MeV and -342$\substack{+130 \\ -111}$ MeV and for PNM priors, we have 22$\substack{+19 \\ -14}$ MeV and -433$\substack{+47 \\ -60}$ MeV.

Overall, a similar level of accuracy of approximately $L$ $\pm 25$ MeV and $K_{\tau}$ $\pm 100$ MeV or better is achieved with a measurement of the neutron skin of lead with an error of $\pm0.06$ fm or a measurement of the neutron skin of Calcium with an error of $\pm0.02$ fm

The results for the PNM priors also starkly reveal the tension between a large value of a neutron skin (if the central value of PREX holds up) and our current state of knowledge of the PNM EOS \cite{Fattoyev:2013aa}. Using the PNM priors, the posterior probability of the radius of lead is 0.21$\substack{+0.02 \\ -0.03}$ fm; the PNM priors do not admit the possibility of obtaining such a high neutron skin. The uninformative priors do so comfortably. It is worth noting that the reason is not only that the uninformative priors have access to higher values of $L$ - values of $L\approx 75$MeV can give sufficiently thick skins - but they access these values of $L$ in regions of parameter space with much smaller symmetry compressibilities than are available to PNM priors.

\begin{table*}[!t]
\caption{\label{tab:table1} 95\% credible ranges (67\% ranges in parentheses) for the symmetry energy parameters using (i) all available neutron skin data combined by taking the highest and lowest bounds reported out of all data points (ii) by adding data in quadrature (iii) using a single hypothetical PREX-II measurement of $\Delta r_{\rm np}^{^{208}\rm Pb}$ = 0.33$\pm0.06$ fm (High), (iv) using a single hypothetical PREX-II measurement of $\Delta r_{\rm np}^{^{208}\rm Pb}$ = 0.15$\pm0.06$ fm (Low) and (v) a single hypothetical CREX measurement of the neutron skin of calcium of $0.18\pm0.02$.}
\begin{ruledtabular}
\setlength{\extrarowheight}{5pt}
\begin{tabular}{ccccc}
 &$J$ (MeV)&$L$ (MeV)&$K_{\rm sym}$(MeV)
 &$K_{\tau}$(MeV) \\
\hline
\multicolumn{5}{l}{
\;\;\; \textbf{Uninformative Priors}}\\
\hline

Full Range& 31.1$\substack{+4.4 (3.2) \\ -5.8 (3.9)}$ & 45$\substack{+35 (20) \\ -33 (20)}$ & -152$\substack{+220 (158) \\ -234 (164)}$ & -426$\substack{+152 (73) \\ -127 (69)}$ \\

Quad& 31.3$\substack{+4.2 (3.1) \\ -5.9 (3.8)}$ & 40$\substack{+34 (21) \\ -26 (16)}$ & -209$\substack{+270 (178) \\ -182 (136)}$ & -444$\substack{+100 (55) \\ -84 (51)}$ \\

PREX-II (High)& 30.6$\substack{+4.8 (3.5) \\ -5.4 (3.8)}$ & 91$\substack{+27 (18) \\ -46 (22)}$ & -199$\substack{+253 (161) \\ -191 (138)}$ & -725$\substack{+249 (115) \\ -208 (111)}$ \\

PREX-II (Low)& 30.6$\substack{+4.9 (3.5) \\ -5.3 (3.7)}$ & 38$\substack{+53 (28) \\ -35 (24)}$ & -125 $\substack{+196 (143) \\ -256 (170)}$ & -384$\substack{+320 (166) \\ -273 (145)}$ \\

CREX & 31.0$\substack{+4.5 (3.3) \\ -5.7 (3.9)}$ & 53$\substack{+46 (25) \\ -43 (25)}$ & -154 $\substack{+222 (159) \\ -232 (162)}$ & -484$\substack{+245 (113) \\ -188 (100)}$\\

\hline
\multicolumn{5}{l}{\;\;\; \textbf{PNM Priors}}\\
\hline

Full Range& 32.0$\substack{+3.0 (1.7) \\ -3.2 (1.7)}$ & 38$\substack{+21 (11) \\ -22 (11)}$ & -256$\substack{+64 (33) \\ -65 (34)}$ & -483$\substack{+68 (36) \\ -64 (36)}$ \\

Quad& 31.9$\substack{+1.3 (0.7) \\ -1.3 (0.7)}$ & 37 $\substack{+9 (4) \\ -8 (4)}$ & -260 $\substack{+35 (19) \\ -33 (19)}$ & -480$\substack{+25 (13) \\ -26 (13)}$ \\

PREX-II (High)& 30.5$\substack{+1.4 (1.1) \\ -5.1 (2.3)}$ & 54$\substack{+14 (9) \\ -34 (15)}$ & -212.9$\substack{+53.3 (33.0) \\ -94.2 (41.4)}$ & -537.5$\substack{+110.9 (49.3) \\ -33.0 (24.0)}$ \\

PREX-II (Low)& 30.7$\substack{+6.1 (3.8) \\ -4.8 (3.3)}$ & 29$\substack{+43 (26) \\ -34 (22)}$ & -279$\substack{+126 (72) \\ -106 (62)}$ & -455$\substack{+104 (71) \\ -133 (83)}$ \\

CREX & 32.7$\substack{+3.2 (2.2) \\ -5.1 (2.7)}$ & 43$\substack{+24 (15) \\ -35 (18)}$ & -243 $\substack{+77 (44) \\ -100 (52)}$ & -500$\substack{+111 (59) \\ -70 (48)}$\\

\end{tabular}
\end{ruledtabular}
\end{table*}

\begin{figure*}[!t]
\includegraphics[scale=0.55]{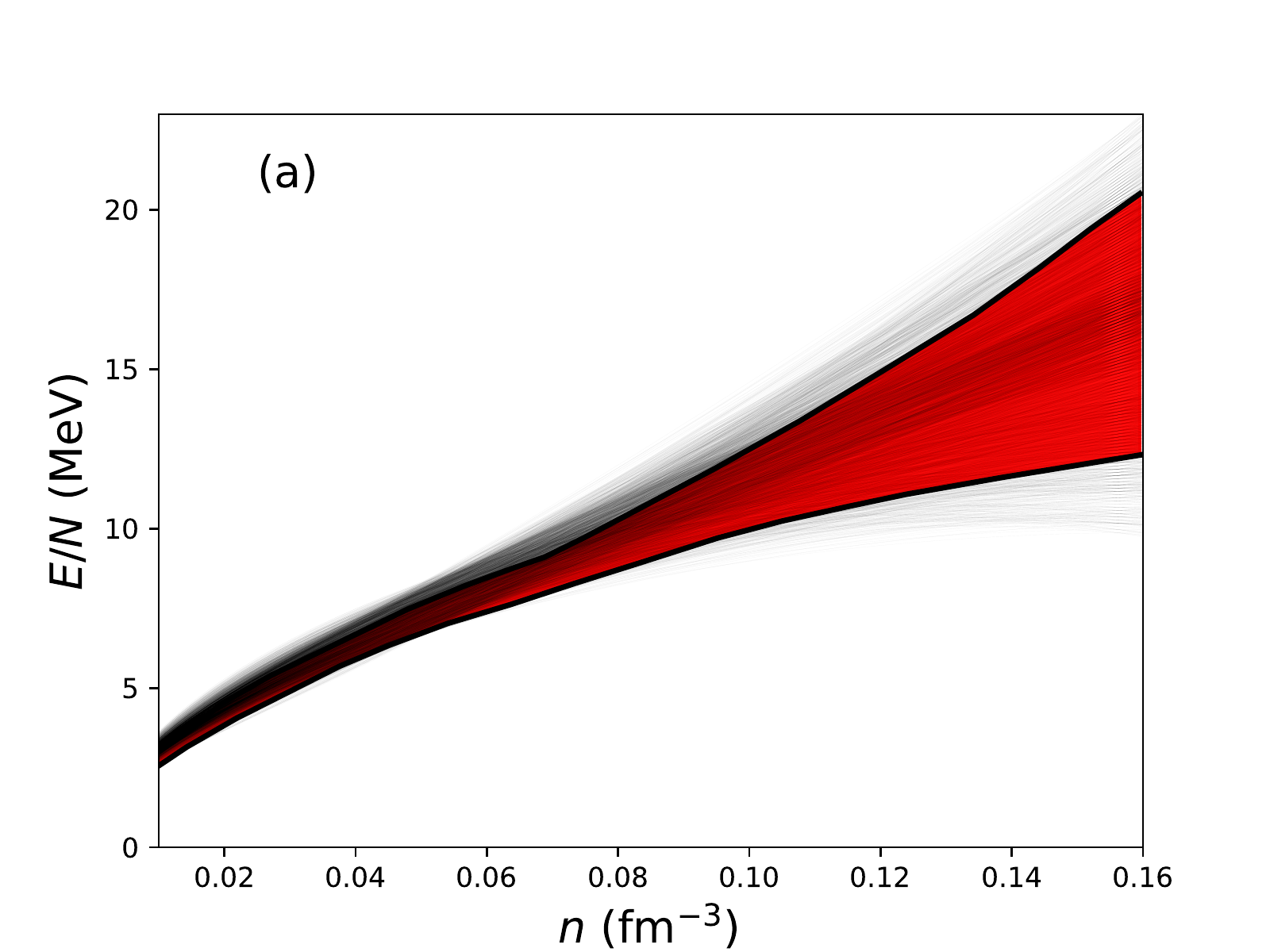}\includegraphics[scale=0.55]{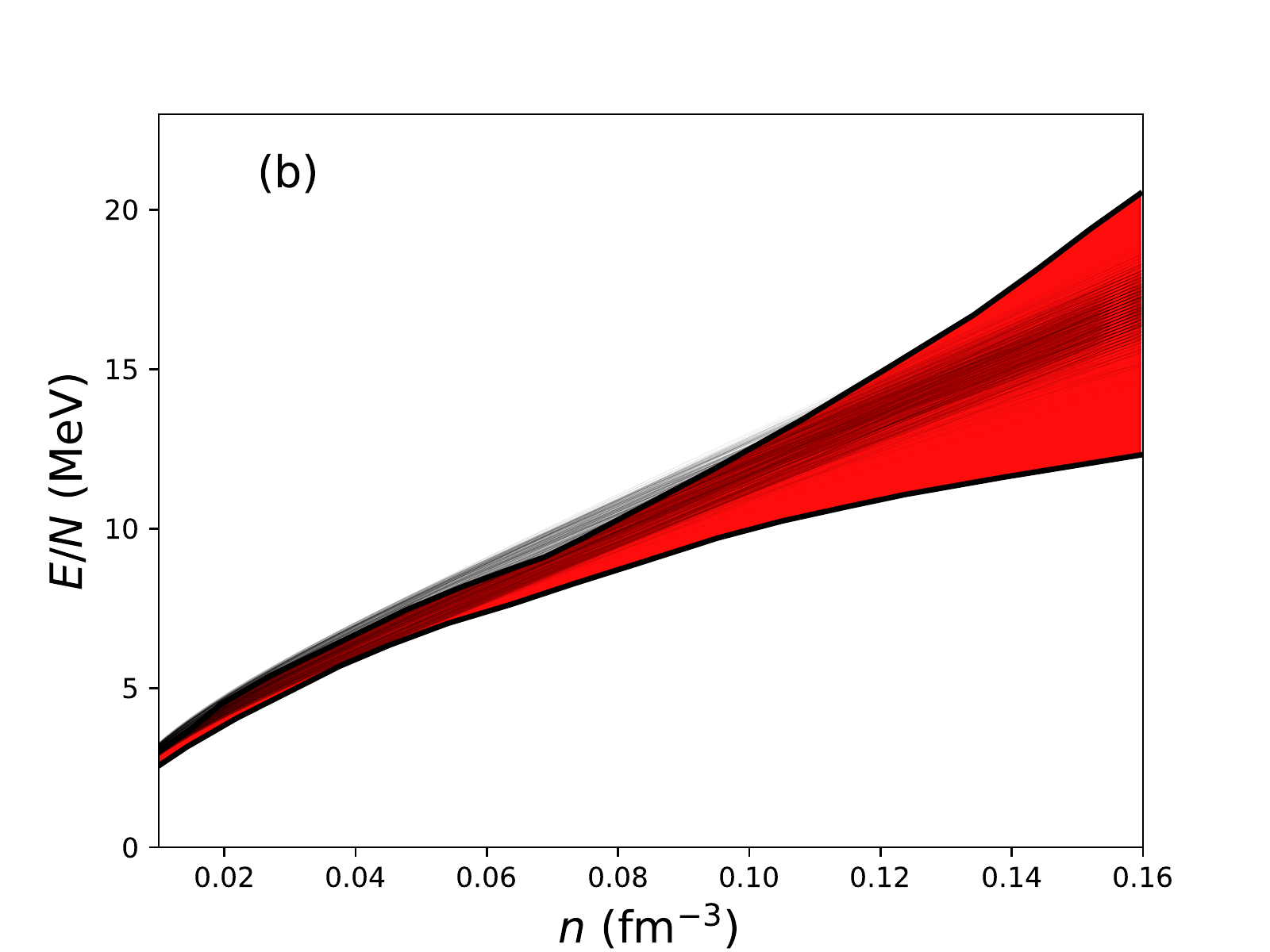}
\caption{Samples of pure neutron matter EOSs drawn from our PNM prior distribution (a) and posterior distribution from neutron skins added in quadrature (b). Neutron skin data significantly constrains the PNM EOS compared to our current state of knowledge.} \label{Fig:11}
\end{figure*}

\section{Discussion and Conclusions}

The results of our Bayesian inference of the first three parameters in the density expansion of the symmetry energy at saturation density, $J$, $L$ and $K_{\rm sym}$, from neutron skin data under two different sets of priors are summarized in Figures~10 and~11 and Table~I. Table~I presents detailed values of the 95\% (67\%) credible ranges for the symmetry energy parameters inferred using the (i) full range of the data for each nuclide (ii) by combining the data for each nuclide in quadrature (iii) a hypothetical large measurement of the neutron skin of lead of $0.33\pm0.06$ from PREX-II, (iv) a hypothetical small measurement of the neutron skin of lead of $0.15\pm0.06$ from PREX-II, and (v) a hypothetical measurement of the neutron skin of calcium of $0.15\pm0.02$ from CREX.

In Figure~10a we compare the 95\% credible regions of $J$ and $L$ obtained with uninformative (blue band) and PNM priors (red circle) with those obtained using a prior to mimic the MSL0 Skyrme model used by \cite{Chen:2010aa} (the orange band) allowing meaningful comparison with the results in \cite{Chen:2010aa}. From the uninformative priors to the MSL0 priors we are moving from a three-parameter family of models ($J$, $L$ and $K_{\rm sym}$) to a two parameter family ($J$ and $L$). The PNM priors represent a three-parameter family ($J$ and the two parameters representing the expansion of the nucleon-nucleon interaction at sub-saturation densities), but those parameters are constrained much more by the theoretical input of chiral-EFT PNM calculations. 

Uninformative priors give a wider band in the $L$-$J$ plane as expected due to the model's extra degree of freedom. Even so, our results show a very similar total range of $L$ values (our inferred range of $L$ for uninformative priors is  14-74 MeV compared to 16-76 MeV with MSL0 priors, and compared with a range of 22-78 MeV in the original analysis \cite{Chen:2010aa}). Our central (median) values of $L\approx$ 40 MeV are significantly lower than the 60 MeV obtained in \cite{Chen:2010aa}. 

It is clear from this figure that including information about PNM makes a large impact on the constraints that can be placed on symmetry energy parameters. The PNM 95\% credible range for $L$ is $29$-$46$ MeV.

Interestingly, a very recent analysis of neutron matter constraints on the symmetry energy parameters with attention to setting rigorous error bounds find a range of $L\approx51-69$ MeV to a $2\sigma$ confidence level \cite{Drischler:2020ab}. This is in the middle of our PNM prior range of 30-90 MeV, but is not consistent with the resulting inference of $L=29$-$46$ MeV from neutron skin data using our PNM prior. Future work will incorporate the results of \cite{Drischler:2020ab} as a prior.

Figure~10b shows the range of inferred values of $L$ versus $K_{\rm sym}$. We show our 95\% credible regions as the blue band (uninformative priors) and red band (PNM priors);  the uninformative priors do not constrain $K_{\rm sym}$ separately, but they do induce a correlation between $L$ and $K_{\rm sym}$. The PNM priors give a 2$\sigma$ result of $K_{\rm sym} = -260^{+35}_{-33}$ MeV. The orange band in Figure~10b shows the region where 95\% of around 500 different nuclear models that are currently in use to describe the properties of terrestrial nuclei  predict values of $L$ and $K_{\rm sym}$ \cite{Tews:2017aa}. Comparing with our inferred region from the PNM prior suggests that the inferred range of $L$ and $K_{\rm sym}$ from neutron skin data and pure neutron matter is inconsistent with models used to predict many other nuclear properties. This suggests the data favors a sub-saturation density dependence of the symmetry energy that is at odds with many existing models. 

Figure~10c shows the range of inferred values of $L$ versus $K_{\tau}$. Previously, neutron skin data was used to obtain simultaneous constraints on $L$ and $K_{\tau}$ within the droplet model \cite{Centelles:2009aa} from whose results we extract the yellow region in the plot. As before, we show our 95\% credible results as the blue region (uninformative priors) and red region (PNM priors), both of which are consistent with the droplet model results (particularly in the range of $K_{\tau}$), but also have substantial areas of their respective posterior distributions at lower values of $L$. A value of $K_{\tau}$=$-500\pm50$ was extracted from analysis of isospin diffusion in heavy ion collisions \cite{Chen:2005aa,Shetty:2007aa} and an analysis of the giant monopole resonance in tin isotopes \cite{Li:2007aa} has led to an inference of $K_{\tau}$=$-550\pm100$. The latter range is indicated by dashed lines in Figure~10c. The PNM priors favor closely the range of $K_{\tau}$=$-500\pm50$ extracted by the isospin diffusion.

Figure~11 shows the effect of the neutron skin data on the pure neutron matter EOS by comparing the EOSs of our priors (left) with those of our posteriors (right). As predicted, \cite{Reinhard:2010aa}, neutron skin data significantly constrains the predicted band of possible pure neutron matter EOSs. 

It is clear that neutron skin data in combination with pure neutron matter calculations provide stronger constraints on the symmetry energy and pure neutron matter EOS than either individually; indeed, we have shown that, starting from a uninformative, wide distribution of possible $J$, $L$ and $K_{\rm sym}$, current neutron skin data alone provides constraints on $L$ that are comparable with, and consistent with current predictions of chiral EFT alone. When taken together, the 95\% credible ranges $L$ and $K_{\tau}$ are reduced by a factor of 4-5. These combined constraints should translate into important limits on the neutron star EOS and, particularly, properties of the neutron star crust, a subject of ongoing investigation.

Finally, we reiterate that more rigorous constraints would be obtained by more direct modeling of the experimental observables - the parity-violating asymmetry, the electric dipole response and the neutron distributions to name but three, using the same sets of nuclear models, and we encourage further work in this direction.

\begin{acknowledgments}
The authors would like to thank Jeremy Holt for useful discussions. WGN and GC acknowledge support from NASA grant 80NSSC18K1019.
\end{acknowledgments}

\end{document}